\documentclass[a4paper,11pt]{article}
\pdfoutput=1 

\usepackage{jcappub,color} 

\usepackage[T1]{fontenc} 

\title{\large{\textcolor{red}{$\;$}\newline\textcolor{magenta}{P.H. Chavanis, S. Kumar, \href{https://doi.org/10.1088/1475-7516/2017/05/018}{JCAP 05 (2017) 018}  [arXiv:1612.01081]}}\newline\textcolor{red}{$\;$}\newline\boldmath \huge{Comparison between the Logotropic and  $\Lambda$CDM models
at the cosmological scale}}

 \author[a]{Pierre-Henri Chavanis,}
 \author[b]{Suresh Kumar}
 \affiliation[a]{Laboratoire de Physique Th\'eorique, Universit\'e Paul
Sabatier,
118 route de Narbonne 31062 Toulouse, France}
\affiliation[b]{Department of Mathematics, BITS Pilani, Pilani Campus,
Rajasthan-333031, India}

\emailAdd{chavanis@irsamc.ups-tlse.fr}
\emailAdd{suresh.kumar@pilani.bits-pilani.ac.in}

\abstract{We perform a detailed comparison between the Logotropic model 
[P.H. Chavanis, Eur. Phys. J. Plus {\bf 130}, 130 (2015)] and the $\Lambda$CDM
model. These two models behave similarly at large (cosmological)
scales up to the present. Differences will appear only in the
far future, in about $25\, {\rm Gyrs}$, when the Logotropic Universe becomes
phantom while the $\Lambda$CDM Universe enters in the de Sitter era. However,
the Logotropic model differs from the $\Lambda$CDM model at small (galactic)
scales, where the latter encounters serious problems. Having a nonvanishing
pressure, the Logotropic model can solve the cusp problem and the missing
satellite problem of the $\Lambda$CDM model. In addition, it leads to dark
matter halos with a constant surface density $\Sigma_0=\rho_0 r_h$, and can
explain its observed value $\Sigma_0=141 \, M_{\odot}/{\rm pc}^2$ without
adjustable parameter. This makes the
logotropic model rather unique among all the models attempting to unify dark
matter and dark energy. In this paper, we compare the Logotropic and
$\Lambda$CDM models at the cosmological scale where they are very close to each
other in order to determine quantitatively how much they
differ. This comparison is facilitated by the fact that these models depend on
only two parameters, the Hubble constant $H_0$ and the present fraction of dark
matter $\Omega_{\rm m0}$. Using the latest observational data from Planck
2015+Lensing+BAO+JLA+HST, we find that the best fit values of $H_0$ and
$\Omega_{\rm m0}$ are $H_0=68.30\, {\rm km}\, {\rm s}^{-1}\,{\rm Mpc}^{-1}$ and
$\Omega_{\rm m0}=0.3014$ for the Logotropic model, and $H_0=68.02\, {\rm km}\,
{\rm s}^{-1}\,{\rm Mpc}^{-1}$ and $\Omega_{\rm m0}=0.3049$ for the $\Lambda$CDM
model. The difference between the two models is at
the percent level. As a result, the Logotropic model competes with the
$\Lambda$CDM model at large scales and solves its problems at small scales.
It may therefore represent a viable alternative to the $\Lambda$CDM model. Our
study provides an explicit example of a theoretically motivated model that is
almost  indistinguishable from the  $\Lambda$CDM model at the present time while
having a completely different (phantom) evolution in the future. We
analytically derive the statefinders of the Logotropic model for
all values of the logotropic constant $B$. We show that the parameter $s_0$
is directly related to this constant since $s_0=-B/(B+1)$ independently of any
other parameter like $H_0$ or $\Omega_{\rm m0}$. For the predicted value of
$B=3.53\times 10^{-3}$, we obtain
$(q_0,r_0,s_0)=(-0.5516,1.011,-0.003518)$ instead of
$(q_0,r_0,s_0)=(-0.5427,1,0)$ for the $\Lambda$CDM model corresponding to
$B=0$.}

\begin{document}
\maketitle
\flushbottom

\section{Introduction}

The nature of dark matter
(DM) and dark energy (DE) is still unknown and remains one of the greatest
mysteries of modern cosmology. DM has been introduced in
astrophysics to
account for the missing mass of the galaxies inferred from the virial theorem
\cite{zwicky}. The existence
of DM has been confirmed by more precise observations of
rotation curves \cite{flat}, gravitational lensing \cite{massey}, and hot gas in
clusters \cite{b2}. DE has been introduced in cosmology to account for the
current acceleration of the expansion of the Universe revealed by the high
redshift of type Ia supernovae
treated as standardized candles \cite{novae}. Recent observations of
baryonic acoustic oscillations (BAO) \cite{b12},   cosmic microwave
background (CMB) anisotropy, microlensing, and the statistics of quasars and
clusters
provide another independent support to the DE hypothesis.
In both cases (DM and DE) more indirect measurements come
from the CMB and large scale structure
observations \cite{smoot,jarosik,planck2013,planck2015}. 
In the standard
cold dark matter ($\Lambda$CDM) model, DM is assumed to be made of weakly
interacting massive particles (WIMPs) with a mass in the
GeV-TeV range. They may correspond to
supersymmetric (SUSY) particles \cite{susy}. These particles freeze out from
thermal
equilibrium in the
early Universe and, as a consequence of this decoupling,  cool off rapidly as
the Universe expands. As a result, DM can be represented by a pressureless
fluid. On the other hand, in the $\Lambda$CDM model, DE is ascribed to the
cosmological constant $\Lambda$ introduced by Einstein
\cite{einsteincosmo}. This is the simplest way to account for the acceleration
of the Universe. The $\Lambda$CDM model
works remarkably well at large (cosmological) scales and is consistent with ever
improving measurements of the CMB from WMAP and
Planck missions \cite{planck2013,planck2015}. However, it encounters serious
problems at small (galactic) scales. In particular, it predicts that DM halos
should be cuspy \cite{nfw} while observations reveal that they have a flat 
core \cite{observations}. On the other hand, the $\Lambda$CDM model predicts
an over-abundance of small-scale structures (subhalos/satellites), much more
than what is observed around the Milky Way \cite{satellites}. These problems are
referred to as the cusp problem \cite{cusp} and  missing satellite problem
\cite{satellites}. The
expression ``small-scale crisis of CDM'' has been coined.

On the other hand, at the cosmological scale, despite its success at
explaining many observations, the $\Lambda$CDM model has to face two theoretical
problems. The first one is the cosmic
coincidence problem, namely why the ratio of DE and DM is of order
unity today if they are two different entities \cite{ccp}. The second one is
the
cosmological constant problem \cite{weinbergcosmo}. The
cosmological constant $\Lambda$
is equivalent to a constant energy density
$\epsilon_{\Lambda}=\rho_{\Lambda}c^2=\Lambda c^2/8\pi
G$ associated with an equation of state $P=-\epsilon$
involving a negative
pressure. Some authors \cite{vacuum} have proposed to
interpret
the  cosmological constant in terms of the vacuum energy.
Cosmological
observations lead to the value
$\rho_{\Lambda}=\Lambda/8\pi G=6.72\times 10^{-24}\,
{\rm g}\, {\rm m}^{-3}$ of the cosmological density (DE).  However, particle
physics and quantum field theory
predict that the vacuum energy should be of the order of the Planck
density $\rho_P=c^5/\hbar G^2=5.16\times 10^{99}\, {\rm g}\, {\rm m}^{-3}$. The
ratio between
the Planck density $\rho_P$ and the cosmological density $\rho_{\Lambda}$ is
\begin{equation}
\frac{\rho_P}{\rho_{\Lambda}}\sim 10^{123},
\label{l1}
\end{equation}
so these quantities differ by $123$ orders of magnitude! This is the origin of
the cosmological constant problem.

In order to remedy these difficulties, several alternative models of DM and DE
have been introduced. The small scale problems of the $\Lambda$CDM model are
related to the assumption that DM is pressureless. Therefore, some authors have
considered the possibility of warm  DM \cite{wdm}. Other authors have
proposed to take into account
the quantum nature of the particles which can give rise to a pressure even at
$T=0$. For example, the DM particle could be a fermion, like the sterile
neutrino, with a mass in the keV
range \cite{vega,clm}. In the fermionic scenario, gravitational collapse is
prevented by the quantum pressure arising from the Pauli exclusion principle. 
 Alternatively, the DM particle could be a boson, like the QCD  axion, with a
mass of the order of $10^{-4}\, {\rm eV}$. Other types of axions with a mass
ranging from $10^{-2}\, {\rm eV}$ to $10^{-22}\, {\rm eV}$ (ultralight
axions) have also been proposed
\cite{marsh}. At $T=0$, the
bosons
form
Bose-Einstein condensates (BECs) so that DM halos could be gigantic
self-gravitating BECs (see the reviews 
\cite{revueabril,revueshapiro,bookspringer}).\footnote{QCD
axions with a mass  $m\sim 10^{-4}\, {\rm eV}$ 
interaction can form mini axion stars with a
maximum mass $M_{\rm max}\sim 10^{-13}\, M_\odot$ \cite{bectcoll} that could be
the
constituents  of DM halos in the form of mini-MACHOS. Ultralight
axions with a mass $m\sim 10^{-22}\, {\rm eV}$ can form DM halos with a
typical mass $M\sim 10^{8}\,
M_\odot$ \cite{revueabril,revueshapiro,bookspringer}.} The bosons
may be noninteracting (fuzzy) \cite{fuzzy,witten} or
self-interacting \cite{si1,si2}. They are equivalent to a scalar field that can
be
interpreted as the wavefunction of the condensate. In the bosonic
scenario,  gravitational collapse is
prevented by the quantum pressure arising from the Heisenberg uncertainty
principle or from the the scattering of the bosons. In the fermionic and
bosonic models, the quantum pressure prevents gravitational collapse and leads
to cores instead of cusps. Quantum mechanics may therefore be a way to solve the
CDM
small-scale crisis.

The physical nature of DE is more uncertain and a  plethora of theoretical
models has been introduced to account for the observation of an accelerating
Universe. Some authors have proposed to abandon the cosmological
constant $\Lambda$ and explain the acceleration of the Universe in terms of DE
with a time-dependent density. The simplest class of
models are those with a constant equation of state parameter $w=P/\epsilon$
called ``quiessence'' (the acceleration of the Universe requires $w<-1/3$). The
case of a time-dependent  equation of state parameter $w(t)$ is called
``kinessence''. Examples of kinessence include scalar fields such as
``quintessence'' \cite{quintessence} and tachyons \cite{tachyon}, as well as
braneworld models of DE \cite{dgp00,deffayet,bw1} and Galileon gravity
\cite{gal10}. Due to the strange properties of DE, the Universe may be phantom
(the energy density increases with time)
\cite{caldwell} possibly giving rise to a big rip \cite{caldwell03} or
a little rip
\cite{littlerip}.

There has been some attempts to introduce models that unify DM and DE. A famous
model is the Chaplygin gas which is an exotic fluid characterized by an 
equation of
state $P=-A/\epsilon$ \cite{chap01}. It behaves as a pressureless
fluid (DM) at early times, and as a
fluid with a constant energy density  (DE) at late times, yielding an
exponential acceleration of the Universe similar to the effect of the
cosmological constant.
However, in the intermediate regime of interest, this model does not give a good
agreement with the observations \cite{sandvik} so that various extensions
of the Chaplygin gas model have been considered, called the generalized
Chaplygin gas \cite{gcg1,gcg2,gcg3} or the polytropic gas
\cite{poly1,poly2,poly3,quadratic}.

Recently, one of us (P.H.C) has introduced another model attempting to unify DM
and DE.
This is the so-called  Logotropic model \cite{delong,decourt}. The Logotropic
model has the following
nice features. At large (cosmological)
scales, the
Logotropic model is almost indistinguishable from the $\Lambda$CDM model up to
the present. They
will differ in about 25 Gyrs years when the Logotropic model becomes
phantom (the energy density increases with time) while the $\Lambda$CDM model
enters in a de Sitter stage (the energy density tends towards a
constant). The fact that the  Logotropic model is almost indistinguishable
from  the $\Lambda$CDM model  at the present time is
nice because the $\Lambda$CDM model is remarkably successful to account
for the large-scale structure of the Universe.
However, the Logotropic model differs from the $\Lambda$CDM model at small
(galactic) scales and is able to solve many problems of the   $\Lambda$CDM
model:

(i) The Logotropic model has a nonvanishing pressure, a nonzero  speed of
sound and a nonzero Jeans length, unlike the CDM model. The pressure can
prevent gravitational collapse and solve the cusp problem and the
missing satellite problem of the CDM model.

(ii) When applied to DM halos, the Logotropic model yields a universal rotation
curve that coincides, up to the halo radius $r_h$, with the empirical Burkert
profile that fits a lot of observational rotation curves
\cite{observations}.\footnote{At larger distances, the halos appear to be more
confined than
predicted by the Logotropic model, a feature which may be explained  by
complicated physical processes such as incomplete
relaxation, evaporation, stochastic forcing from the external environment etc.
As a result, the density profiles of the halos decrease at large distances as
$r^{-3}$ like the NFW \cite{nfw} and Burkert \cite{observations} profiles
instead of $r^{-1}$ as predicted by the Logotropic model.} In particular, the
Logotropic density profile presents a core like the Burkert \cite{observations}
profile while the CDM density profile presents a cusp \cite{nfw} .

(iii) The Logotropic model explains the universality of the surface
density $\Sigma_0=\rho_0 r_h$  of DM halos \cite{donato}, the universality of
the mass $M_{300}$ of dwarf
spheroidal galaxies (dSphs)  contained within a sphere of size $r_u=300\, {\rm
pc}$ \cite{strigari}, and the Tully-Fisher relation $v_h^4\propto M_b$
\cite{tf}. It predicts, without free parameter,  the numerical value of
$\Sigma_0=\rho_0 r_h=141 \, M_{\odot}/{\rm pc}^2$, $M_{300}=1.93\times 10^7\,
M_{\odot}$ and $M_b/v_h^4=44\, M_{\odot} {\rm km}^{-4} {\rm s}^4$. These
theoretical
predictions agree remarkably well with the observations \cite{delong}.

These nice properties make the Logotropic model rather unique
among all the models attempting to unify DM and DE. It is therefore important to
compare the Logotropic and $\Lambda$CDM models at the cosmological scale in
order to
determine how close they are. This comparison is
interesting because the Logotropic model is completely different from the
$\Lambda$CDM model on a theoretical point of view. Therefore, it is important to
quantify precisely their difference, even small.  We stress that, unlike many
other
theoretical models, the Logotropic model has no adjustable parameter so that it
is
fully predictive. More precisely, it only depends on two fundamental
parameters, the Hubble constant $H_0$ and the present fraction of
DM $\Omega_{\rm m0}$, like the  $\Lambda$CDM model. This allows us to make a
very
accurate comparison between the two models. We find that the difference
between the two models is at
the percent level which is beyond observational precision. Therefore, the 
Logotropic model may be a viable
alternative to the $\Lambda$CDM model: it
competes with the
$\Lambda$CDM model at large scales where the $\Lambda$CDM model works well
and solves its problems at small scales. On the other hand, our
study provides an explicit example of a theoretically motivated model that is
almost  indistinguishable from the  $\Lambda$CDM model at present while having a
completely different (phantom) evolution in the future.

The paper is organized as follows. Section \ref{sec_lc} summarizes the theory of
\cite{delong,decourt} with a new presentation and complements. Section
\ref{sec_stl} analytically derives the statefinders of the Logotropic model and
provides their typical values. Section \ref{sec_fine} compares the Logotropic
and $\Lambda$CDM models in the light of the latest observational data
from {\it Planck} 2015+Lensing+BAO+JLA+HST.  Section
\ref{sec_conclusion} concludes. Readers who are familiar
with the Logotropic model \cite{delong,decourt}, or who are only interested 
in the comparison with the observations, may directly go to Sec. 
\ref{sec_fine}.

{\it Remark:} Throughout the paper, we provide
general equations that are valid for arbitrary values of the Logotropic
constant $B$. Interestingly, we show that the statefinder parameter $s_0$ is
directly related to the Logotropic constant $B$ since $s_0=-B/(B+1)$
independently of any
other parameter like $H_0$ or $\Omega_{\rm m0}$. This can
be useful to parameterize deviations between the Logotropic model and the
$\Lambda$CDM
model (or other models) in situations where
the parameter $B$ is large. This can also be useful to constrain the value of
this parameter from cosmological observations.  Indeed, a large value of $B$
leads to statefinders that substantially differ from the $\Lambda$CDM model.
However, in the numerical
applications and in the figures, we take the
value
$B=3.53\times 10^{-3}$ predicted by the theory \cite{delong,decourt}.

\section{Logotropic cosmology}
\label{sec_lc}

\subsection{Unification of dark matter and dark energy by a single dark fluid}
\label{sec_df}

We assume that the Universe is homogeneous and isotropic, and contains a uniform
perfect fluid of energy density $\epsilon(t)$, rest-mass density $\rho(t)$, and
isotropic pressure $P(t)$. It will be called the dark fluid (DF).  We assume
that the
Universe is flat ($k=0$) in agreement with the observations of the CMB
\cite{planck2013,planck2015}.  On the other
hand,  we ignore the
cosmological constant ($\Lambda=0$) because the contribution of DE
will be taken into account in the equation of state of the DF. Under
these
assumptions, the Friedmann equations can be written as \cite{weinbergbook}:
\begin{equation}
\frac{d\epsilon}{dt}+3\frac{\dot a}{a}(\epsilon+P)=0,
\label{df1}
\end{equation}
\begin{equation}
\label{df2}
\frac{\ddot a}{a}=-\frac{4\pi G}{3c^2} \left (\epsilon+3P\right ),
\end{equation}
\begin{equation}
H^2=\left (\frac{\dot a}{a}\right )^2=\frac{8\pi
G}{3c^2}\epsilon,
\label{df3}
\end{equation}
where $a(t)$ is the scale factor and $H=\dot a/a$ is the Hubble parameter.
Among these equations, only two are independent. The first equation is the
equation of continuity, or the energy conservation equation. The second equation
determines the acceleration of the Universe. The third equation
relates the Hubble parameter, i.e., the velocity of expansion of the Universe, to
the energy density. The deceleration parameter
is defined by
\begin{equation}
\label{df4}
q(t)=-\frac{{\ddot a}a}{{\dot a}^2}.
\end{equation}
The Universe is decelerating when $q>0$ and accelerating when $q<0$. 
Introducing the equation of state parameter $w=P/\epsilon$, and using the
Friedmann equations (\ref{df2}) and (\ref{df3}), we obtain for a flat Universe
\begin{equation}
\label{df5}
q(t)=\frac{1+3w(t)}{2}.
\end{equation}
We see from Eq. (\ref{df5}) that the Universe is decelerating if $w>-1/3$
(strong energy condition) and accelerating if $w<-1/3$.\footnote{According to
general relativity, the source for the
gravitational potential is $\epsilon+3P$. Indeed, the spatial part ${\bf g}$
of
the geodesic acceleration satisfies the exact equation $\nabla\cdot {\bf
g}=-4\pi G (\epsilon+3P)$ showing that the source of geodesic acceleration is
$\epsilon+3P$ not $\epsilon$ \cite{paddycosmo}. Therefore, in general
relativity, gravitation becomes ``repulsive'' when $P<-\epsilon/3$.} On the
other hand,
according to Eq. (\ref{df1}), the energy density decreases with the scale
factor if
$w>-1$ (null dominant energy condition) and increases with the scale factor if
$w<-1$. The latter case corresponds to a ``phantom'' Universe \cite{caldwell}.

The local form of the first law of thermodynamics can be expressed
as \cite{weinbergbook}:
\begin{equation}
d\left (\frac{\epsilon}{\rho}\right )=-P d\left (\frac{1}{\rho}\right )+T d\left
(\frac{s}{\rho}\right ),
\label{df6}
\end{equation}
where $\rho=n m$ is the mass density, $n$ is the number density, and $s$ is
the
entropy density in the rest frame. For a relativistic fluid at $T=0$, or for an
adiabatic evolution such that $d(s/\rho)=0$ (which is the
case for a perfect fluid), the first law of thermodynamics
reduces to
\begin{equation}
d\epsilon=\frac{P+\epsilon}{\rho}d\rho.
\label{df7}
\end{equation}
For a given equation of state, Eq. (\ref{df7}) can be integrated to obtain the
relation between the energy density $\epsilon$ and the rest-mass density $\rho$.
If the equation of state is prescribed under the form $P=P(\rho)$, Eq.
(\ref{df7}) can be written as a first order linear differential equation:
\begin{equation}
\frac{d\epsilon}{d\rho}-\frac{1}{\rho}\epsilon=\frac{P(\rho)}{\rho}.
\label{df8}
\end{equation}
Using the method of the variation of the constant, we obtain \cite{delong}:
\begin{equation}
\epsilon=\rho c^2+\rho\int^{\rho}\frac{P(\rho')}{{\rho'}^2}\, d\rho'=\rho
c^2+u(\rho),
\label{df9}
\end{equation}
where the constant of integration is determined in such a way that the function
$u(\rho)$ does not contain any contribution linear in $\rho$. We note that
$u(\rho)$ can be interpreted as an internal energy density
\cite{delong} (see also Appendix \ref{sec_gtet}). Therefore, the
energy density $\epsilon$ is the sum of the rest-mass energy $\rho c^2$ and the
internal energy $u(\rho)$. The rest-mass energy is positive while the internal
energy can be  positive or negative. Of course, the total
energy $\epsilon=\rho
c^2+u(\rho)$ is always positive.

Combining the first law of thermodynamics (\ref{df7})  with
the continuity equation (\ref{df1}), we get \cite{delong}:
\begin{equation}
\frac{d\rho}{dt}+3\frac{\dot a}{a}\rho=0.
\label{df10}
\end{equation}
We note that this equation is exact for a fluid at $T=0$, or for a perfect
fluid, and that it does not depend on the explicit form of the equation of state
$P(\rho)$. It expresses the conservation of the rest-mass.
It can be integrated into
\begin{equation}
\rho=\frac{\rho_0}{a^3},
\label{df11}
\end{equation}
where $\rho_0$ is the present value of the rest-mass density of the DF, and the
present
value of the scale factor is taken to be $a_0=1$. 

The previous results suggest the following interpretation \cite{delong}. The
energy density of the DF
\begin{equation}
\epsilon=\rho c^2+\rho\int^{\rho}\frac{P(\rho')}{{\rho'}^2}\, d\rho'=\rho
c^2+u(\rho)=\frac{\rho_0c^2}{a^3}+u\left (\frac{\rho_0}{a^3}\right
)=\epsilon_{\rm m}+\epsilon_{\rm new}
\label{df12}
\end{equation}
is the sum of two terms: a rest-mass energy term $\rho c^2\propto a^{-3}$ that
mimics DM and an internal energy term $u(\rho)$ that mimics a
``new fluid''.
This ``new fluid'' may have different meanings depending on the
equation of state $P(\rho)$ as discussed in the Appendix of \cite{stiff}. For
an equation of state
$P=-\epsilon_{\Lambda}$, where $\epsilon_{\Lambda}$ is a constant (cosmological
density), we find that
\begin{equation}
\epsilon=\rho c^2+u(\rho)=\rho
c^2+\epsilon_{\Lambda}=\frac{\rho_0c^2}{a^3}+\epsilon_{\Lambda},
\label{df13}
\end{equation}
which is equivalent to the $\Lambda$CDM model. In that case, the ``new fluid''
is equivalent to the cosmological constant or to DE with a constant density.
More
generally, when the
equation of state is close to a negative constant, the ``new fluid'' describes
DE with a time-dependent density \cite{delong}.

\subsection{The Logotropic dark fluid}
\label{sec_ldf}

Following \cite{delong}, we assume that the Universe is filled with a
single DF
described by a
Logotropic equation of state\footnote{The logotropic equation of state was
introduced phenomenologically in
astrophysics
by McLaughlin and  Pudritz \cite{pud} to describe the internal
structure and the
average properties of molecular clouds and clumps. It was also studied by
Chavanis and Sire \cite{logo} in the context of Tsallis generalized
thermodynamics \cite{tsallis} where it was shown to
correspond to a
polytropic equation of state of the form $P=K\rho^{\gamma}$ with
$\gamma\rightarrow 0$ and $K\rightarrow \infty$ in such a way
that $A=\gamma K$ is finite. In Appendix \ref{sec_gtet}, we develop this analogy
with
generalized thermodynamics.}
\begin{equation}
P=A\ln\left (\frac{\rho}{\rho_P}\right ),
\label{ldf1}
\end{equation}
where $A$ is a constant with the dimension of an energy density that is
called the Logotropic temperature (see Appendix \ref{sec_gtet}) and $\rho_P$ is
a constant with the
dimension of a mass density. These constants will be determined in Sec.
\ref{sec_lt}. The
fluid described by the equation of state (\ref{ldf1}) is called the Logotropic
Dark Fluid
(LDF). Using Eqs. (\ref{df9}) and (\ref{ldf1}), the relation
between the energy density and the rest-mass density is \cite{delong}:
\begin{equation}
\epsilon=\rho c^2+u(\rho)=\rho c^2-A\ln \left (\frac{\rho}{\rho_P}\right )-A.
\label{ldf2}
\end{equation}
The energy density is the sum of two terms: a rest-mass energy term $\rho
c^2$ that mimics DM and
an internal energy term $u(\rho)$ that mimics DE. This decomposition
leads to a natural, and physical, unification of DM and DE and elucidates their
mysterious nature \cite{delong}.  In the present approach, we
have a single DF. However, in order to
make the connection with the traditional approach where the Universe is assumed
to
be composed of DM and DE, we identify the rest-mass energy of the
DF with the energy density of DM\footnote{For convenience, we also include the
contribution of the baryons in the rest-mass energy of the dark
fluid so that $\epsilon_{\rm m}$ represents the total energy density of matter
(baryonic matter $+$ DM). In principle, the DF and the baryonic fluid
must be treated as two separate species. However, since the final equations are
the same, we find it more economical to group them together from the
start.}
\begin{equation}
\epsilon_{\rm m}=\rho c^2=\frac{\rho_0c^2}{a^3}
\label{ldf3}
\end{equation}
and we identify the internal energy of the DF with the energy density of DE
\begin{equation}
\epsilon_{\rm de}=u=-A\ln \left (\frac{\rho}{\rho_P}\right )-A=-A\ln \left
(\frac{\rho_0}{\rho_Pa^3}\right )-A.
\label{ldf4}
\end{equation}
The pressure is related to the internal energy, or to the energy density of the
DE, by the affine equation
of state $P=-u-A=-\epsilon_{\rm de}-A$. We note that the internal
energy (DE density) is  positive for $\rho<\rho_P/e$ and negative for
$\rho>\rho_P/e$. In the present approach, having $\epsilon_{\rm de}<0$  is possible
since, as
we have explained,  $\epsilon_{\rm de}$ does not really correspond to DE but to the
internal energy $u$ of the DF.\footnote{Note that the Logotropic model that
attempts to
unify DM and DE is only valid at sufficiently late times where the density is
low. Therefore, $\epsilon_{\rm de}$ is always positive in practice.} Combining Eqs.
(\ref{ldf1}) and
(\ref{ldf2}), we obtain  
\begin{equation}
\epsilon=\rho_P c^2 e^{P/A}-P-A
\label{ldf5}
\end{equation}
which determines, by inversion, the equation of state $P(\epsilon)$ of the
LDF \cite{delong}. Combining  Eqs. (\ref{df11}), (\ref{ldf1}) and (\ref{ldf2}),
we get 
\begin{equation}
P=A\ln\left (\frac{\rho_0}{\rho_P a^3}\right )
\label{ldf6}
\end{equation} 
and
\begin{equation}
\epsilon=\frac{\rho_0 c^2}{a^3}-A\ln\left (
\frac{\rho_0}{\rho_Pa^3}\right ) -A,
\label{ldf7}
\end{equation}
which give the evolution of the pressure and energy density of the
LDF as a
function of the scale factor. The LDF is normal (the energy density decreases
with the scale factor) for $a<a_M$ and phantom (the energy density increases
with the scale factor) for $a>a_M$, where $a_M=(\rho_0c^2/A)^{1/3}$. At that
point,  the energy density reaches its minimum value $\epsilon_M=-A\ln
({A}/{\rho_P c^2})$. We have $\rho_M={A}/{c^2}$ and $P_M=-\epsilon_M$. We
note that $A/c^2$ is equal to the rest-mass density of the LDF at the point
where
it becomes phantom.

In the early Universe ($a\rightarrow 0$, $\rho\rightarrow +\infty$),  the
rest-mass energy (DM) dominates, so that
\begin{equation}
\epsilon\sim \rho c^2\sim \frac{\rho_0 c^2}{a^3},\qquad P\sim A\ln \left
(\frac{\epsilon}{\rho_P c^2}\right ).
\label{ldf8}
\end{equation}
We emphasize that the pressure of the LDF is nonzero, even in the early
Universe. However, since $P\ll\epsilon$ for $a\rightarrow
0$, everything happens in the Friedmann equations (\ref{df1})-(\ref{df3}) {\it
as if} the fluid were pressureless ($P\simeq 0$). Therefore, for small values of
the scale factor, we obtain $\epsilon\propto a^{-3}$ as in the CDM model
($P=0$).\footnote{Since the
Friedmann equations  (\ref{df1})-(\ref{df3}) govern the large scale structure of
the Universe (the
cosmological background), we conclude that pressure effects are negligible at
large scales in the early Universe. However, at small scales, corresponding to
the size of DM halos, pressure effects encapsulated
in the Logotropic equation of state (\ref{ldf1}) become important and can solve
the problems
of the CDM
model such as the cusp problem and the missing satellite problem as
shown
in \cite{delong,decourt}.  }

In
the late Universe ($a\rightarrow +\infty$, $\rho\rightarrow 0$), the internal
energy (DE) dominates, and we have
\begin{equation}
\epsilon\sim
-A\ln \left (\frac{\rho}{\rho_P}\right )\sim 3A \ln a,\qquad P\sim -\epsilon.
\label{ldf9}
\end{equation}
We note that the equation
of state $P(\epsilon)$ of the LDF behaves asymptotically as
$P\sim -\epsilon$, similarly to the usual equation of state of DE. It is
interesting to recover the equation of state $P=-\epsilon$ from the Logotropic
model (\ref{ldf1}). This was not obvious {\it a priori}. More precisely,
if we keep the constant terms in the asymptotic formulae (because of the slow
change of the logarithm), we obtain 
\begin{equation}
\epsilon\simeq
-A\ln \left (\frac{\rho}{\rho_P}\right )-A \simeq
-A\ln \left (\frac{\rho_0}{\rho_Pa^3}\right )-A,\qquad  P\simeq -\epsilon-A.
\label{ldf10}
\end{equation}

\subsection{The general equations}
\label{sec_be}

The Logotropic model depends on three unknown parameters $A$, $\rho_P$ and
$\rho_0$. Applying Eqs.
(\ref{ldf3}) and (\ref{ldf4})  at $a=1$, 
we obtain the identities
\begin{equation}
\epsilon_{\rm m0}=\Omega_{\rm m0}\epsilon_0=\rho_0 c^2,
\label{lt1}
\end{equation}
\begin{equation}
\epsilon_{\rm de0}=\Omega_{\rm de0}\epsilon_0=u_0=-A\ln \left
(\frac{\rho_0}{\rho_P}\right )-A,
\label{lt2}
\end{equation}
where $\epsilon_0={3H_0^2c^2}/{8\pi
G}$ is the
present energy density of the Universe, 
$\Omega_{\rm m0}$ is the present fraction of DM (rest mass of the
DF),\footnote{As explained in footnote 5, $\Omega_{\rm m0}$
represents the present fraction of
(baryonic $+$ dark) matter.} and
$\Omega_{\rm de0}=1-\Omega_{\rm m0}$ is the present fraction of DE
(internal energy of the DF). We
write
the Logotropic temperature as
\begin{equation}
A=B \Omega_{\rm de0}\epsilon_0,
\label{lt3i}
\end{equation}
where $B$ is the dimensionless Logotropic temperature. From
Eqs.
(\ref{lt1})-(\ref{lt3i}),
we
obtain
\begin{equation}
B=\frac{1}{\ln\left
(\frac{\rho_P}{\rho_0}\right )-1}=\frac{1}{\ln\left
(\frac{\rho_Pc^2}{\Omega_{\rm m0}\epsilon_0}\right )-1}.
\label{lt4}
\end{equation}
Using the above relations, we see that our initial set of unknown parameters
$(A,\rho_P,\rho_0)$ is equivalent to  $(H_0,\Omega_{\rm m0},B)$. After simple
manipulations, the general equations giving the normalized rest-mass density,
pressure and energy density of the LDF can be
expressed in terms of $B$ as 
\begin{equation}
\frac{\rho
c^2}{\epsilon_0}=\frac{\Omega_{\rm m0}}{a^3},
\label{be1}
\end{equation}
\begin{equation}
\label{be2}
\frac{P}{\Omega_{\rm de0}\epsilon_{0}}
=-B-1+B\ln\left (\frac{\rho
c^2}{\epsilon_0\Omega_{\rm m0}}\right ),
\end{equation}
\begin{equation}
\frac{P}{\Omega_{\rm de0}\epsilon_{0}}=-B-1-3B\ln a,
\label{be3}
\end{equation}
\begin{equation}
\frac{\epsilon}{\epsilon_0}=\frac{\rho
c^2}{\epsilon_0}+\Omega_{\rm de0}\left\lbrack 1-B\ln\left
(\frac{\rho c^2}{\Omega_{\rm m0}\epsilon_0}\right )\right\rbrack,
\label{be4}
\end{equation}
\begin{equation}
\frac{\epsilon}{\epsilon_0}=\frac{\Omega_{\rm m0}}{a^3}+\Omega_{\rm de0}(1+3B\ln
a),
\label{be5}
\end{equation}
\begin{equation}
\frac{\epsilon}{\epsilon_0}=\Omega_{\rm m0}e^{(B+1)/B}e^{P/B\Omega_{\rm de0}\epsilon_{0}
}
-\Omega_{\rm de0}\left (\frac{P}{\Omega_{\rm de0}\epsilon_{0}}+B\right ),
\label{be6}
\end{equation}
\begin{equation}
w=\frac{-\Omega_{\rm de0}(B+1+3B\ln
a)}{\frac{\Omega_{\rm m0}}{a^3}+\Omega_{\rm de0}(1+3B\ln a)}.
\label{be6w}
\end{equation}
In the early Universe, we obtain
\begin{equation}
\frac{\epsilon}{\epsilon_0}\sim\frac{\rho
c^2}{\epsilon_0}\sim\frac{\Omega_{\rm m0}}{a^3},
\label{be7}
\end{equation}
\begin{equation}
\frac{P}{\Omega_{\rm de0}\epsilon_{0}}\simeq -B-1+B\ln\left
(\frac{\epsilon}{\Omega_{\rm m0}\epsilon_{0}}\right ),
\label{be8}
\end{equation}
\begin{equation}
w\sim -(B+1+3B\ln a)\frac{\Omega_{\rm de0}}{\Omega_{\rm m0}}a^3.
\label{be8w}
\end{equation}
In the
late Universe, we get
\begin{equation}
\frac{\epsilon}{\epsilon_0}\simeq\Omega_{\rm de0}\left\lbrack 1-B\ln\left
(\frac{\rho c^2}{\Omega_{\rm m0}\epsilon_0}\right
)\right\rbrack\simeq\Omega_{\rm de0}(1+3B\ln
a),
\label{be9}
\end{equation}
\begin{equation}
\frac{P}{\Omega_{\rm de0}\epsilon_{0}}\simeq
-B-\frac{\epsilon}{\Omega_{\rm de0}\epsilon_{0}},
\label{be10}
\end{equation}
\begin{equation}
w\simeq -1-\frac{B}{1+3B\ln a}.
\label{be10w}
\end{equation}

\subsection{The $\Lambda$CDM model ($B=0$)}
\label{sec_lcdm}

The $\Lambda$CDM model is
recovered for $B=0$. In that case, Eqs.
(\ref{be1})-(\ref{be6w}) reduce to
\begin{equation}
\frac{\rho
c^2}{\epsilon_0}=\frac{\Omega_{\rm m0}}{a^3},\qquad
\frac{P}{\Omega_{\rm de0}\epsilon_{0}} =-1,
\label{be11}
\end{equation}
\begin{equation}
\frac{\epsilon}{\epsilon_0}=\frac{\rho
c^2}{\epsilon_0}+\Omega_{\rm de0},\qquad
\frac{\epsilon}{\epsilon_0}=\frac{\Omega_{m0 } } { a^3 }+\Omega_{\rm de0},
\label{be13}
\end{equation}
\begin{equation}
w=\frac{-\Omega_{\rm de0}}{\frac{\Omega_{\rm m0}}{a^3}+\Omega_{\rm de0}}.
\label{be6wl}
\end{equation}
In the early Universe, we obtain
\begin{equation}
\frac{\epsilon}{\epsilon_0}\sim\frac{\rho
c^2}{\epsilon_0}\sim\frac{\Omega_{\rm m0}}{a^3},\qquad w\sim
-\frac{\Omega_{\rm de0}}{\Omega_{\rm m0}}a^3.
\label{be15}
\end{equation}
In the late Universe, we get
\begin{equation}
\frac{\epsilon}{\epsilon_0}\simeq\Omega_{\rm de0},\qquad w\rightarrow -1.
\label{be16}
\end{equation}
The  $\Lambda$CDM model depends on two unknown parameters $H_0$ and
$\Omega_{\rm m0}$. In the $\Lambda$CDM model, DM is given by
$\epsilon_{\rm m}=\Omega_{\rm m0}\epsilon_0/a^3$ and DE is 
constant: $\epsilon_{\rm de}=\epsilon_{\Lambda}=\Omega_{\rm de0}
\epsilon_0$. The
$\Lambda$CDM model is equivalent to a single DF with a
constant negative pressure
$P=-\epsilon_{\Lambda}$ leading to the relation
$\epsilon=\rho
c^2+\epsilon_{\Lambda}=\rho_0
c^2/a^3+\epsilon_{\Lambda}=\epsilon_{\rm m0}/a^3+\epsilon_{\Lambda}$ between the
energy
density $\epsilon$ and the
rest-mass density $\rho$ or scale factor $a$ \cite{delong}.

\subsection{The Logotropic model ($B=3.53\times
10^{-3}$)}
\label{sec_lt}

It is convenient to
introduce the notation
$\epsilon_{\Lambda}=\rho_{\Lambda}c^2=\epsilon_{\rm de0}=\Omega_{\rm de0}
\epsilon_0=(1-\Omega_{\rm m0})\epsilon_0$. In the $\Lambda$CDM model,
$\epsilon_\Lambda=\Lambda c^2/8\pi G$
represents the constant value of DE. More generally, $\epsilon_\Lambda$
represents the
present value of DE. It will be called the cosmological
density.\footnote{Since $\Omega_{\rm de0}\sim \Omega_{\rm m0}\sim 1$, the present DE
density $\epsilon_\Lambda=\epsilon_{\rm de0}$ is of the same order of magnitude as
the present DM energy density $\epsilon_{\rm m0}$ or as the present energy density
of the Universe $\epsilon_0$. This observation is refered to as the cosmic
coincidence problem \cite{ccp}. Since, in the Logotropic model, DM and
DE are two manifestations  of the same DF (representing its rest mass energy and
internal energy), the cosmic coincidence problem may be alleviated
\cite{delong}.} With
this
notation,
the Logotropic temperature can be written as
\begin{equation}
A=B \epsilon_\Lambda \qquad {\rm with}\qquad B=\frac{1}{\ln\left
(\frac{1-\Omega_{\rm m0}}{\Omega_{\rm m0}}\frac{\rho_P}{\rho_{\Lambda}}\right )-1}.
\label{lt3}
\end{equation}
The second relation of Eq. (\ref{lt3}) can be rewritten as
\begin{equation}
\frac{\rho_P}{\rho_{\Lambda}}=\frac{\Omega_{\rm m0}}{1-\Omega_{\rm m0}}e^{1+1/B}.
\label{l10}
\end{equation}
As observed in \cite{delong}, this identity is
strikingly similar to Eq. (\ref{l1}) which
appears in relation to the
cosmological constant problem. {\it Inspired by this analogy, \cite{delong}
postulated that $\rho_P$ is the Planck
density $\rho_P=c^5/G^2\hbar=5.16\times 10^{99}\, {\rm
g}\, {\rm m}^{-3}$.}\footnote{{At the begining of the study made in 
\cite{delong}, the reference density in the Logotropic equation of state
(\ref{ldf1}) was
unspecified, and denoted $\rho_*$. The dimensionless Logotropic parameter $B$
was treated as a free parameter related to $\rho_*$. When applied to DM
halos, the Logotropic equation of state was found to generate density profiles
with a constant surface density $\Sigma_0=\rho_0 r_h=5.8458... (A/4\pi
G)^{1/2}$ provided that $A$ is treated as a universal constant. It was remarked
that this result is in agreement with observations that show that the surface
density of DM halos is constant \cite{donato}. By comparing the observational
value $\Sigma_0=141 \, M_{\odot}/{\rm pc}^2$ with the theoretical one
$\Sigma_0=5.8458... (A/4\pi
G)^{1/2}$, it was found that $B=A/\epsilon_\Lambda$ is equal to
$3.53\times 10^{-3}$ implying
that $\rho_*$ is
huge,  of the order of the Planck density
$\rho_P=5.16\times 10^{99}\, {\rm
g}\, {\rm m}^{-3}$. As a result, it was proposed in \cite{delong} to
{\it identify} $\rho_*$ with
$\rho_P$. It was then proceeded the other way round. If we postulate from the
start that
$\rho_*=\rho_P$, we find that
$B$ is {\it determined} by Eq. (\ref{lt3}) yielding $B=3.53\times 10^{-3}$.
We then obtain $\Sigma_0=141 \, M_{\odot}/{\rm pc}^2$ in remarkable
agreement with the observations. In
parallel, it was observed in \cite{delong} that the identity (\ref{l10}) is
analogous to Eq.
(\ref{l1}) giving further support to the choice of identifying $\rho_*$ with the
Planck density $\rho_P$.}
} In that case, the
identity (\ref{lt3}) determines the dimensionless Logotropic temperature $B$.
Approximately, $B\simeq
1/\ln(\rho_P/\rho_{\Lambda})\simeq 1/[123\ln(10)]$. This relation gives a
new interpretation to the famous number $123\simeq{\rm
log}(\rho_P/\rho_{\Lambda})$ as being the inverse dimensionless Logotropic
temperature. 

To determine a more precise value of $B$, we substitute
$\epsilon_0={3H_0^2c^2}/{8\pi G}$ and $\rho_P=c^5/G^2\hbar$ into Eq.
(\ref{lt3}). This
gives
\begin{equation}
B=\frac{1}{\ln\left
(\frac{8\pi c^5}{3G\hbar\Omega_{\rm m0}H_0^2}\right
)-1}.
\label{lt5}
\end{equation}
This equation shows that $B$ is determined by fundamental constants such as $c$,
$G$ and $\hbar$, and by the cosmological parameters $\Omega_{\rm m0}$ and $H_0$. 
Therefore, there are only
two unknown parameters in the Logotropic
model,  $\Omega_{\rm m0}$ and $H_0$, like in the $\Lambda$CDM model. In
addition, the value of $B$ is rather insensitive to the exact values of
$\Omega_{\rm m0}$ and $H_0$ because these quantities appear in a logarithm.
This allows us to treat $B$ as a fundamental constant \cite{delong}. To see
that, we rewrite Eq. (\ref{lt5}) under the form 
\begin{equation}
B=\frac{1}{290.135-\ln(\Omega_{\rm m0})-2\ln\left (\frac{H_0}{\text{km s}^{-1}
\text{Mpc}^{-1}}\right )}.
\label{lt6}
\end{equation}
We can estimate $B$ by taking the values of $\Omega_{\rm m0}$ and $H_0$
obtained from the $\Lambda$CDM model. The values of the cosmological
parameters adopted in \cite{delong} (not the most updated ones) are
$\Omega_{\rm m0}=0.274$, $\Omega_{\rm de0}=1-\Omega_{\rm m0}=0.726$, 
$H_0=70.2 \, {\rm km}\,  {\rm s}^{-1}\, {\rm Mpc}^{-1}=2.275\, 10^{-18} \,
{\rm s}^{-1}$, $\epsilon_0/c^2={3H_0^2}/{8\pi
G}=9.26\times 10^{-24}\, {\rm
g}\, {\rm m}^{-3}$, $\epsilon_{\rm m0}/c^2=\Omega_{\rm m0}\epsilon_0/c^2=2.54\times
10^{-24}\,
{\rm g}\, {\rm
m}^{-3}$, and $\epsilon_{\rm de0}/c^2=\Omega_{\rm de0}\epsilon_0/c^2=6.72\times
10^{-24}\, {\rm g}\, {\rm m}^{-3}$.  With
these values, we get \cite{delong}:
\begin{equation}
B=3.53\times 10^{-3},\qquad A=2.13\times 10^{-9}
\, {\rm g}\, {\rm m}^{-1}\, {\rm s}^{-2}. 
\label{lt7}
\end{equation}
The important point is that the value of $B$ is rather insensitive to
the precise values  of $\Omega_{\rm m0}$ and $H_0$. Even if we make an error of one
order of magnitude (!) on the values of $\Omega_{\rm m0}$ and $H_0$ (while these
values
are known with a high precision), we get almost the
same value of $B$. Therefore, the value   of $B$ given in Eq. (\ref{lt7}) is
fully reliable and we shall adopt it in the following. Using updated values of 
$\Omega_{\rm m0}$ and $H_0$ in Sec. \ref{sec_fine}, we show that the value of $B$ is
not changed. In conclusion, there is no free parameter in the
Logotropic
model. From now on, we shall regard $A$ and $B$ as fundamental constants that
supersede the cosmological constant $\Lambda$. We note that they depend on
all the fundamental constants of physics $\hbar$, $G$, $c$, and
$\Lambda$ [see Eq.  (\ref{lt3})].
Using Eq. (\ref{lt3}), the logotropic equation of state (\ref{ldf1}) can be
rewritten as
\begin{equation}
P=B\rho_{\Lambda}c^2\ln\left (\frac{\rho}{\rho_P}\right ).
\label{add1}
\end{equation}
We note that $P(\rho_{\Lambda})\simeq -\epsilon_{\Lambda}$.

\subsection{Is the Logotropic model a quantum extension of the $\Lambda$CDM
model?}
\label{sec_qe}

We have seen in Sec. \ref{sec_lcdm} that the $\Lambda$CDM model could be
recovered as a limit of the Logotropic model when $B\rightarrow 0$.
According to Eq. (\ref{l10}), the condition $B\rightarrow 0$ is equivalent to
$\rho_P\rightarrow 
+\infty$, hence $\hbar\rightarrow 0$. Therefore, the $\Lambda$CDM model
appears, in the approach of \cite{delong}, as a semi-classical approximation of
the Logotropic model corresponding to $\hbar\rightarrow 0$. If the Planck
constant were strictly equal to zero ($\hbar=0$), we would have $B=0$ and the
$\Lambda$CDM
model would be obtained. However, since the Planck
constant is small but nonzero ($\hbar\neq 0$), the parameter $B$ has a small but
nonzero value given by Eq. (\ref{lt7}). This  leads to a model  {\it different}
from the $\Lambda$CDM model. The constant $B$ has a quantum nature
since it depends on $\hbar$ [see Eq. (\ref{lt5})]. The fact that the nonzero
value of $B$ predicted by
the
Logotropic model is confirmed by the observations (see Ref. \cite{delong})
shows  that quantum mechanics ($\hbar\neq 0$) plays a role in cosmology in
relation to DM and DE. This may suggest a link with a theory of quantum
gravity. In other words, we may wonder whether the Logotropic model can be
interpreted as a quantum extension of the $\Lambda$CDM
model. The precise meaning to give to this statement remains, however, to be
established.

\subsection{The evolution of the Logotropic Universe}
\label{sec_es}

The evolution of the Logotropic Universe has been described in
detail in \cite{delong,decourt}. Here, we simply summarize the main results. 
In the Logotropic model, using Eq. (\ref{be5}), the Friedmann equation
(\ref{df3}) takes
the form
\begin{equation}
H=\frac{\dot a}{a}=H_0 \sqrt{\frac{\Omega_{\rm m0}}{a^3}+\Omega_{\rm de0}(1+3B\ln
a)}.
\label{es1}
\end{equation}
The temporal evolution of the scale factor $a(t)$ is given by
\begin{equation}
\int_0^a
\frac{dx}{x\sqrt{\frac{\Omega_{\rm m0}}{x^3}+\Omega_{\rm de0}(1+3B\ln
x)}}=H_0 t.
\label{es2}
\end{equation}
For $B=0$, corresponding to the $\Lambda$CDM model, Eq. (\ref{es2}) can be
integrated
analytically leading to the well-known solution
\begin{equation}
a=\left (\frac{\Omega_{\rm m0}}{\Omega_{\rm de0}}\right
)^{1/3}\sinh^{2/3}\left (\frac{3}{2}\sqrt{\Omega_{\rm de0}}H_0 t\right
),\qquad \frac{\epsilon}{\epsilon_0}=\frac{\Omega_{\rm de0}}{\tanh^2 \left
(\frac{3}{2}\sqrt{\Omega_{\rm de0}}H_0 t\right )}.
\label{dmde6}
\end{equation}
For $B\neq 0$, Eq. (\ref{es2}) must be integrated numerically. However, its
asymptotic
behaviors can be obtained analytically.

For $t\rightarrow 0$, we can neglect the contribution of DE ($\Omega_{\rm de0}=0$)
and we
obtain
\begin{equation}
a\sim \left (\frac{3}{2}\sqrt{\Omega_{\rm m0}}H_0 t\right )^{2/3},\qquad
\frac{\epsilon}{\epsilon_0}\sim \frac{4}{9H_0^2 t^2}.
\label{es3}
\end{equation}
This coincides with the Einstein-de Sitter (EdS) solution originally obtained
for a pressureless Universe ($P=0$); see footnote 7. In this asymptotic
regime, the results are independent of $B$. Therefore, Eq. (\ref{es3}) is valid
both for the  $\Lambda$CDM model ($B=0$) and for the Logotropic model
($B\neq 0$).

For $t\rightarrow +\infty$, we can neglect the contribution of
DM ($\Omega_{\rm m0}=0$). For the
$\Lambda$CDM model ($B=0$), we obtain the de Sitter (dS) solution
\begin{equation}
a\sim \left (\frac{\Omega_{\rm m0}}{4\Omega_{\rm de0}}\right
)^{1/3}e^{\sqrt{\Omega_{\rm de0}}H_0t},\qquad \epsilon\simeq \epsilon_\Lambda.
\end{equation}
The Hubble parameter tends towards a constant:
\begin{equation}
\frac{H}{H_0}\rightarrow
\sqrt{\Omega_{\rm de0}}.
\end{equation}
Numerically, $H\rightarrow 1.94\times 10^{-18}\, {\rm s}^{-1}$. For the
Logotropic model
($B\neq 0$), we
find
\begin{equation}
a\propto e^{\frac{3B}{4}\Omega_{\rm de0}H_0^2t^2},\qquad
\frac{\epsilon}{\epsilon_0}\sim
\left (\frac{3B}{2}\Omega_{\rm de0}H_0 t\right )^2.
\label{es3b}
\end{equation}
The energy density increases with time meaning that the Universe is
phantom. The scale factor has a
super de Sitter behavior represented by a stretched exponential
\cite{delong,decourt}.
The Hubble parameter increases linearly with time and its time derivative tends
towards a constant:
\begin{equation}
\frac{H}{H_0}\sim \frac{3B}{2}\Omega_{\rm de0}H_0 t, \qquad \dot H\rightarrow 
\frac{3B}{2}\Omega_{\rm de0}H_0^2.
\label{es3d}
\end{equation}
For the Logotropic model with $B=3.53\times 10^{-3}$, we get $\dot
H\rightarrow 1.99\times 10^{-38}\, {\rm s}^{-2}$. We note that the preceding
equations can be expressed in terms of $A$ according to
\begin{equation}
a\propto e^{\frac{2\pi G A}{c^2}t^2},\qquad
\epsilon\sim \frac{6\pi G}{c^2}A^2t^2,\qquad H\sim \frac{4\pi
G A}{c^2}t,\qquad \dot
H\rightarrow  \frac{4\pi
G A}{c^2}.
\label{es3c}
\end{equation}

\begin{center}
\begin{figure}[htb]\centering
\includegraphics[width=8cm]{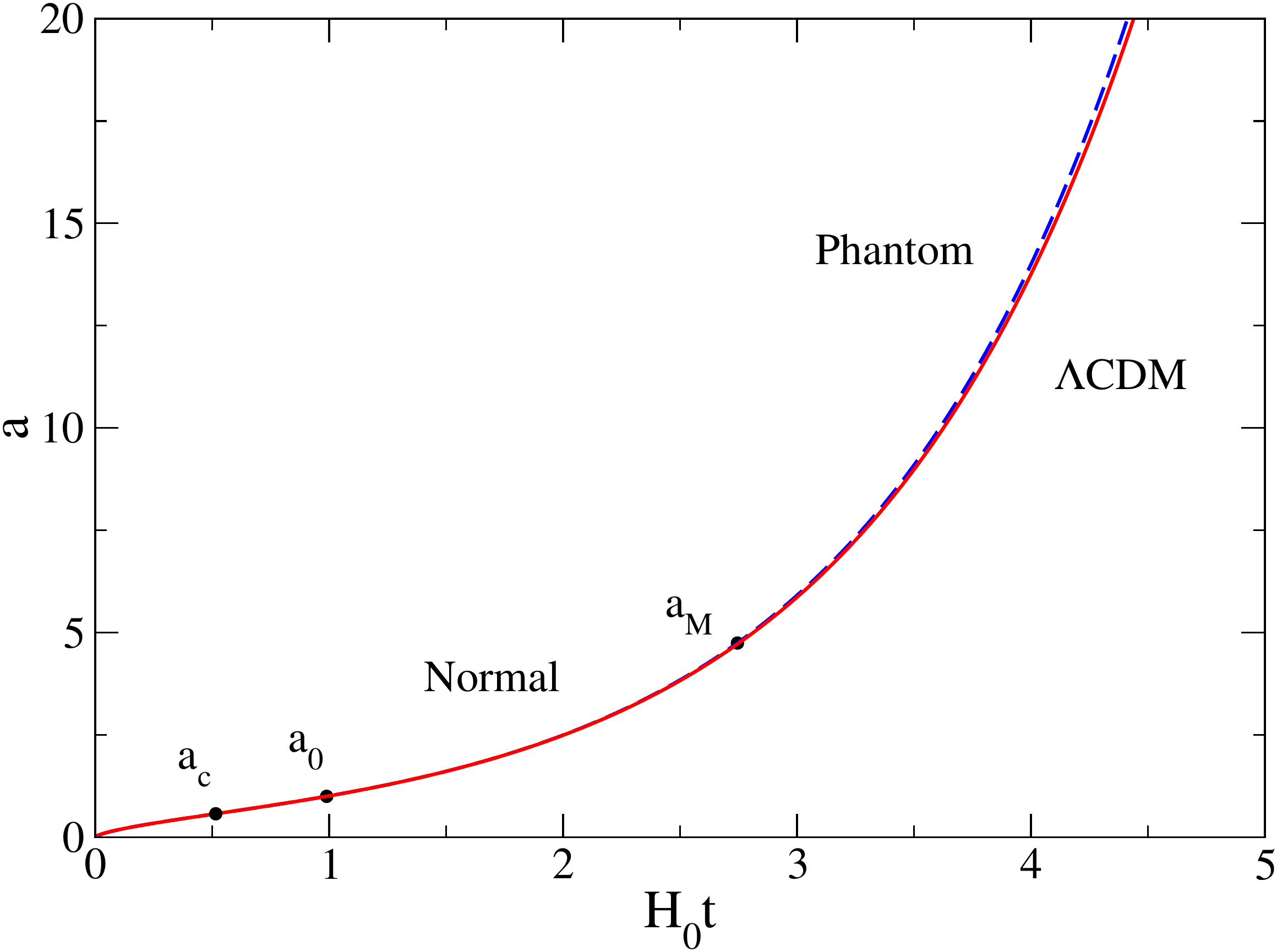}
\caption{Temporal evolution of the
scale factor in the Logotropic model (blue) as compared to the $\Lambda$CDM
model (red).}\label{talettreK}
\end{figure}
\end{center}

\begin{center}
\begin{figure}[htb]\centering
\includegraphics[width=8cm]{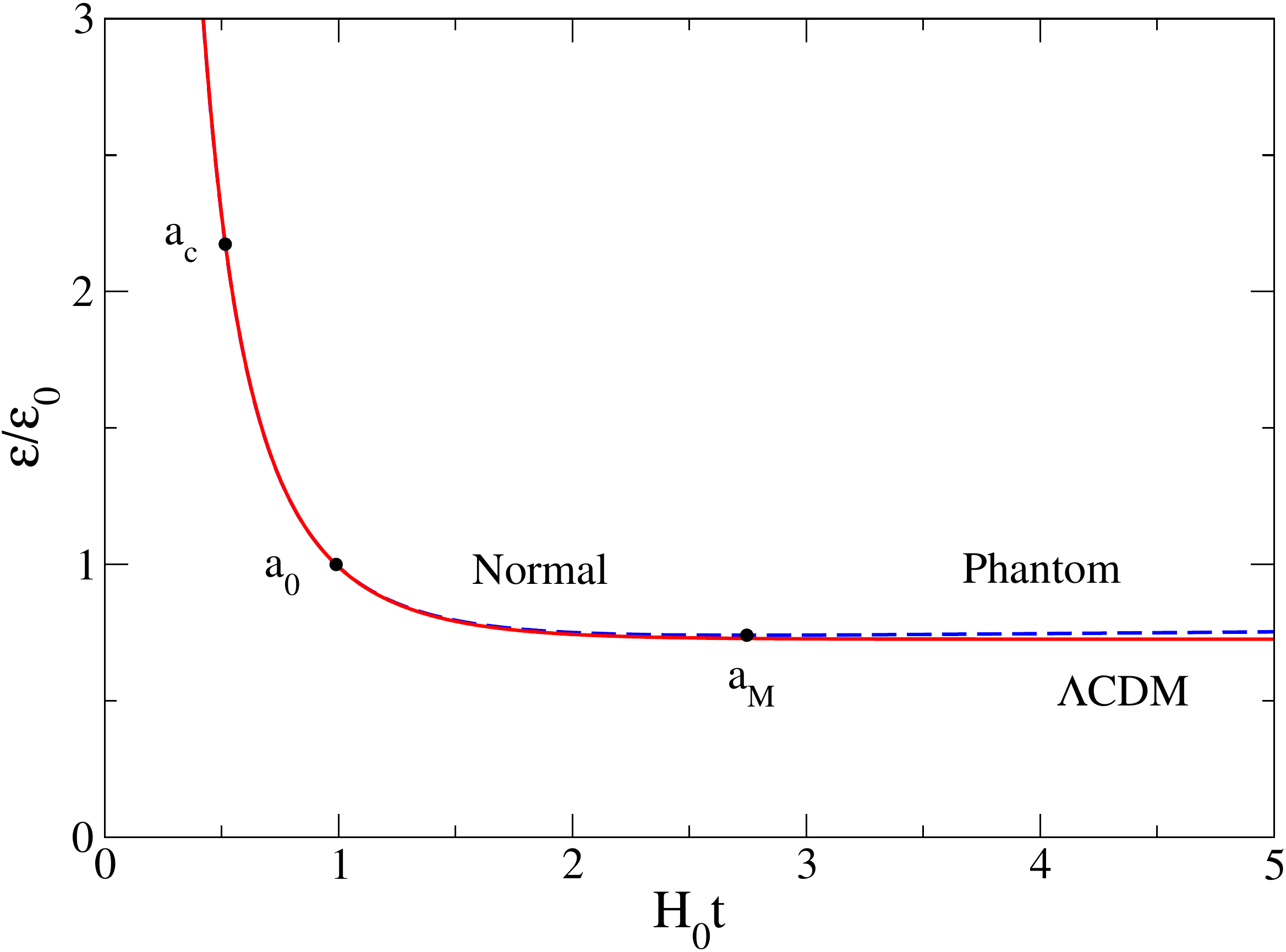}
\caption{Temporal evolution of the energy density in the Logotropic model (blue)
 as
compared to the $\Lambda$CDM
model (red).}\label{teps}
\end{figure}
\end{center}

\begin{center}
\begin{figure}[htb]\centering
\includegraphics[width=8cm]{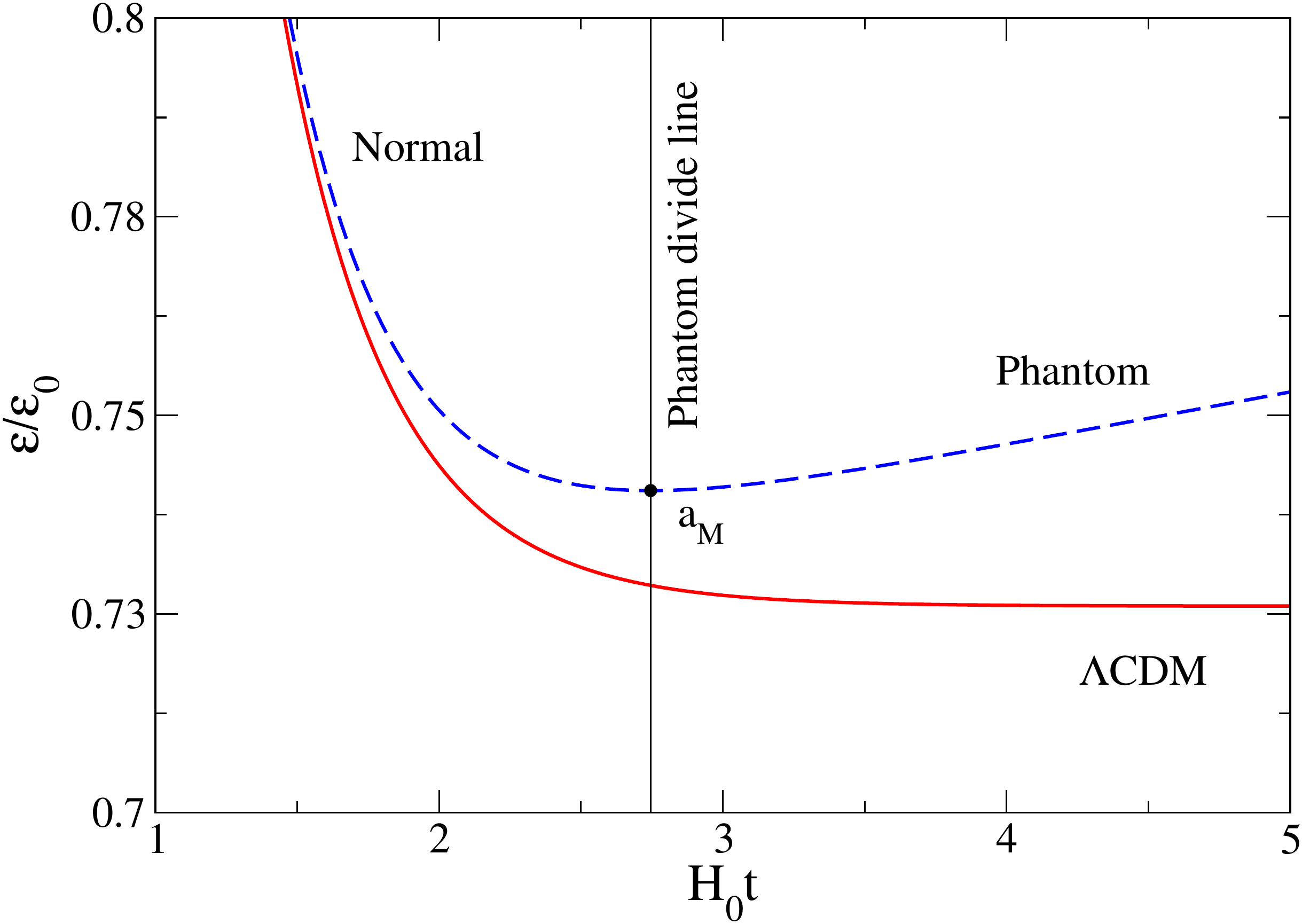}
\caption{Zoom of Fig. \ref{teps}. We
have indicated
the phantom divide line at which the energy density starts increasing with
the
scale factor in the Logotropic model.}\label{tepszoom}
\end{figure}
\end{center}

The temporal evolutions of the energy density $\epsilon(t)$ and of the scale
factor $a(t)$ are represented in Figs. \ref{talettreK}-\ref{tepszoom}. We have
taken $B=3.53\times
10^{-3}$. The
Universe starts at $t=0$ with a vanishing
scale factor ($a=0$) and an infinite energy density
($\epsilon\rightarrow +\infty$).\footnote{Of course, the Logotropic model that
attempts to unify DM and DE is only valid at sufficiently late times. If we want
to describe the very early Universe, we must take into account the
inflation and radiation eras. Therefore, the limit $a\rightarrow 0$ is here
formal. We note that $u=\epsilon_{\rm dm}$ becomes positive for $a>a_*$ with
$a_*=e^{-1/3B}=9.775\times
10^{-42}$.} The Universe experiences a DM era followed by a DE era.  In the
DM era, the Universe is decelerating. The scale factor increases
as $t^{2/3}$ and the energy density  decreases as $t^{-2}$. This corresponds to
the EdS solution. In the DE era, the Universe is accelerating. The Universe
starts accelerating at $t_c=7.19\, {\rm Gyrs}$ (corresponding to $a_c=0.574$ and
$\epsilon_c/\epsilon_0=2.17$). The
energy density $\epsilon_{\rm m}$ associated with DM (actually the rest-mass energy of
the
DF) is equal to the energy density $\epsilon_{\rm de}$ associated with DE
(actually the internal energy of the DF) at $t_2=9.63\, {\rm Gyrs}$
(corresponding to $a_2=0.723$ and $\epsilon_2/\epsilon_0=1.45$). The
Logotropic
model is  very close to the $\Lambda$CDM model up to the
present (the age of the Universe is $t_0=13.8\, {\rm Gyrs}$).
However, in
the far future, at $t_M=38.3\, {\rm Gyrs}$
(corresponding to $a_M=4.75$ and $\epsilon_M/\epsilon_0=0.7405$), the
Logotropic Universe will become phantom. At that
moment, the energy will increase with time as the
Universe expands. Asymptotically, its energy density will increase as $t^2$ and
the
scale factor will have a super de Sitter behavior. The scale factor and the
energy
density will become infinite in infinite time. This corresponds to a little rip
\cite{littlerip}. By contrast, in the $\Lambda$CDM model, the energy density of
the Universe tends towards a constant $\epsilon_{\Lambda}$ and the scale factor
has a
de Sitter behavior. 

{\it Remark:} The Logotropic model
may break down before the Universe enters in the phantom regime because the
speed of
sound exceeds the speed of light at $t_s=34.5\, {\rm Gyrs}$ (corresponding
to $a_s=3.77$ and $\epsilon_s/\epsilon_0=0.741$), i.e., before the
Universe
becomes phantom ($a_s=a_M/2^{1/3}$). Note that the speed of sound $c_s$ defined
by
$c_s^2=P'(\epsilon)c^2=c^2/((a_M/a)^3-1)$ is real for
$a<a_M$ (i.e., when the Universe is normal) and imaginary for $a>a_M$ (i.e., when
the Universe is phantom). We must remain cautious, however, about these
considerations because it has been known for a long time that the propagation
of signals with a speed bigger than the speed of light is possible and does
not contradict the general principles of physics \cite{gm}.

\subsection{The two fluids model}
\label{sec_tf}

In the Logotropic model developed in  \cite{delong} and discussed previously,
the Universe is made of a single DF with an equation of state given by Eq.
(\ref{ldf1}) unifying DM and DE.  It is interesting to consider a related model
in
which the Universe is made of
two noninteracting fluids, a DM fluid with a pressureless equation of state
\begin{equation}
P_{\rm m}=0, \qquad \frac{\epsilon_{\rm m}}{\epsilon_0}=\frac{\Omega_{\rm m0}}{a^3},\qquad
\epsilon_{\rm m}=\rho_{\rm m} c^2,
\label{tf1}
\end{equation}
and a DE fluid with an affine equation of state\footnote{This
equation of state is studied in Appendix A of \cite{poly3}.}
\begin{equation}
P_{\rm de}=-\epsilon_{\rm de}-A,\qquad 
\frac{\epsilon_{\rm de}}{\epsilon_0}=\Omega_{\rm de0}(1+3B\ln a),\qquad
\epsilon_{\rm de}=-A\ln\rho_{\rm de}+C,
\label{tf2}
\end{equation}
where we have defined $B=A/\Omega_{\rm de0}\epsilon_0$ as before.
In
order to
obtain the second and third expressions of each line, we have solved the
equation of continuity (\ref{df1}) and the first law of thermodynamics
(\ref{df8}) for
each individual fluid described by the corresponding equation of state. These
two fluids correspond to the asymptotic behaviors
of the LDF in the early and late Universe respectively. Their equations of state
parameters are  
\begin{equation}
w_{\rm m}=\frac{P_{\rm m}}{\epsilon_{\rm m}}=0,\qquad  
w_{\rm de}=\frac{P_{\rm de}}{\epsilon_{\rm de}}=-1-\frac{B}{1+3B\ln a}.
\label{ha1}
\end{equation}
The function $w_{\rm de}(a)$ starts from $-1$ at
$a_i=0$, increases, tends
towards $+\infty$ as $a\rightarrow a_*^{-}=9.775\times 10^{-42}$, tends  
towards
$-\infty$ as $a\rightarrow a_*^{+}=9.775\times 10^{-42}$, increases and tends
slowly  (logarithmically) towards
$-1^{-}$ for $a\rightarrow
+\infty$. For typical values of $a$, the parameter $w_{\rm de}$ has an
approximately constant value $\sim w_{\rm de0}=-1-B=-1.00353$, due to
its slow
(logarithmic) dependence on the scale factor.

Summing the energy
contribution of these two fluids,
we obtain
\begin{equation}
\frac{\epsilon}{\epsilon_0}=\frac{\Omega_{\rm m0}}{a^3}+\Omega_{\rm de0}(1+3B\ln
a)
\label{tf3}
\end{equation}
which coincides with Eq.  (\ref{be5}). The total pressure
$P=P_{\rm m}+P_{\rm de}=P_{\rm de}$ reduces
to the pressure of DE and can be written as
\begin{equation}
\frac{P}{\Omega_{\rm de0}\epsilon_{0}}=-B-1-3B\ln a
\label{tf4}
\end{equation}
which coincides with Eq. (\ref{be3}). Therefore, at the background level, the
two fluids model is equivalent to the single LDF model. However, despite this
equivalence, the one fluid model and the two fluids
model present some differences:

(i) In the two fluids model, the DE fluid exists only for
$a>a_*=e^{-1/3B}=9.775\times
10^{-42}$ because
we must require its energy density to be positive ($\epsilon_{\rm de}\ge 0$). In
the one fluid model, $\epsilon_{\rm de}$ can be negative because it actually
represents the internal energy $u$ of the DF which can be positive or negative
(as long as the total energy $\epsilon=\epsilon_{\rm m}+\epsilon_{\rm de}$ is positive).
However, since $a_*$ is extremely small, corresponding to an epoch where our
study
is not applicable anyway, this difference is not important. 

(ii) In the two fluids model, the pressure, which reduces to the  pressure of DE
is given by $P=A\ln\rho_{\rm de}+C$. Therefore, it depends on the logarithm of the
rest-mass density of DE, $\rho_{\rm de}$, not on the total rest-mass density,
$\rho$, as in the one fluid model. 

(iii) In the two fluids model, there is no way to predict the value of
the constant $B$ while this is possible in the one fluid model (see Sec.
\ref{sec_lt}). 

(iv) Defining the speed of sound by
$(c_s^2/c^2)_i=P'(\epsilon_i)$ for each species $i\in \lbrace {\rm m,de}\rbrace$
in
the two fluids model, we find from Eqs. (\ref{tf1}) and (\ref{tf2}) that  DM
has a vanishing speed of sound $(c_s)_{\rm m}=0$ and that DE has an imaginary
speed of
sound $(c_s^2/c^2)_{\rm de}=-1$.   By contrast, in the one fluid model, defining
the
speed of sound by
$(c_s^2/c^2)=P'(\epsilon)$, the LDF has a real nonzero
speed of sound $(c_s^2/c^2)_{\rm LDF}=1/((a_M/a)^3-1)$ in the normal Universe
($a<a_M=4.75$) \cite{delong}. This difference has several important
consequences:

(iv-a) Even if the one fluid and two fluids models are equivalent at
the background level, they differ at the level of the perturbations.

(iv-b) Since the LDF has a nonzero speed of sound, it has a nonvanishing
Jeans length. This Jeans length may account for the
minimum size of DM
halos in the Universe as discussed in
\cite{delong}. In the two fluids model, DM has a
vanishing speed of sound (like CDM) so there is no minimum size of DM
halos. Therefore, DM halos should form at all scales.
This should lead to an abundance of small-scale structures which are not
observed. This is the so-called missing satellite problem \cite{satellites}. 

(iv-c) The pressure of the LDF can prevent
gravitational collapse and lead to DM halos with a core.  In the two fluids
model, DM has a vanishing pressure (like CDM) 
so that nothing prevents gravitational collapse. This leads to cuspy DM halos
that are in contradiction with observations. This is the so-called cusp
problem \cite{cusp}.

\section{Statefinders of the Logotropic
model}
\label{sec_stl}

\subsection{Definition}

Sahni et al. \cite{sahni03} suggested a very useful way of comparing
and distinguishing different cosmological models by introducing the statefinders
$\{q,r,s\}$ defined by
\begin{equation}
q=-\frac{{\ddot a}a}{{\dot a}^2},\qquad r=\frac{\dddot a}{a H^3},\qquad
s=\frac{r-1}{3(q-1/2)},
\label{stl1}
\end{equation}
where $q$ is the deceleration parameter and $r$ is the jerk parameter.
Introducing the Hubble parameter $H=\dot a/a$,
we obtain
\begin{equation}
q=-1-\frac{aH'}{H},
\label{stl2}
\end{equation}
\begin{equation}
r=a^2\left (\frac{H'}{H}\right )^2+4a\frac{H'}{H}+1+a^2\frac{H''}{H},
\label{stl3}
\end{equation}
where prime denotes a derivative with respect to $a$.

For the Logotropic
model, the Hubble parameter is given by
\begin{equation}
\frac{H}{H_0}=\sqrt{\frac{\Omega_{\rm m0}}{a^3}+\Omega_{\rm de0}(1+3B\ln a)}.
\label{stl4}
\end{equation}
After simplification, we obtain the simple analytical expressions
\begin{equation}
q=\frac{1}{2}-\frac{3\Omega_{\rm de0}}{2}\frac{B+1+3B\ln
a}{\frac{\Omega_{\rm m0}}{a^3}+\Omega_{\rm de0}(1+3B\ln a)},
\label{stl5}
\end{equation}
\begin{equation}
r=1+\frac{9
B\Omega_{\rm de0}}{2}\frac{1}{\frac{\Omega_{\rm m0}}{a^3}+\Omega_{\rm de0}(1+3B\ln
a)},
\label{stl6}
\end{equation}
\begin{equation}
s=-\frac{B}{B+1+3B\ln a}.
\label{stl7}
\end{equation}
We note the remarkable fact that the parameter $s$ is a
universal function of $a$ and $B$ (it does not depend on the
present fraction of DM and DE). We now consider asymptotic limits of these
expressions. For $a\rightarrow 0$:
\begin{equation}
q\simeq \frac{1}{2}-\frac{3\Omega_{\rm de0}}{2\Omega_{\rm m0}}(B+1+3B\ln
a)a^3,
\label{stl8}
\end{equation}
\begin{equation}
r\simeq 1+\frac{9B\Omega_{\rm de0}}{2\Omega_{\rm m0}}a^3.
\label{stl9}
\end{equation}
The Logotropic Universe begins\footnote{See footnote 11.}  at $t_i=0$
corresponding to $a_i=0$ and $\epsilon_i\rightarrow +\infty$ (big
bang).  At that point $q_i=1/2$, $r_i=1$ and $s_i=0$. This corresponds to the
EdS limit.  For $a\rightarrow +\infty$:
\begin{equation}
q\simeq -1-\frac{3B}{2(1+3B\ln a)},
\label{stl10}
\end{equation}
\begin{equation}
r\simeq 1+\frac{9B}{2(1+3B\ln
a)}.
\label{stl11}
\end{equation}
Therefore, $q\rightarrow -1$, $r\rightarrow 1$ and $s\rightarrow 0$. This
corresponds to the super dS limit (little rip).

For the $\Lambda$CDM
model ($B=0$) we recover the well-known expressions
\begin{equation}
q=\frac{1}{2}-\frac{3\Omega_{\rm de0}}{2}\frac{1}{\frac{\Omega_{\rm m0}}{a^3}
+\Omega_{\rm de0}},\qquad r=1,\qquad s=0.
\label{stl12}
\end{equation}
For $a\rightarrow 0$:
\begin{equation}
q\simeq \frac{1}{2}-\frac{3\Omega_{\rm de0}}{2\Omega_{\rm m0}}a^3.
\label{stl13}
\end{equation}
For $a\rightarrow +\infty$:
\begin{equation}
q\rightarrow  -1.
\label{stl14}
\end{equation}
The $\Lambda$CDM model Universe begins at $t_i=0$
corresponding to $a_i=0$ and $\epsilon_i\rightarrow +\infty$ (big
bang).  At that point $q_i=1/2$, $r_i=1$ and $s_i=0$. This corresponds to the
EdS limit. On the other hand, for $a\rightarrow +\infty$ we get  $q\rightarrow
-1$, $r\rightarrow 1$ and $s\rightarrow 0$. This
corresponds to the dS limit. We note that the statefinders of the Logotropic and
$\Lambda$CDM
models coincide for $a\rightarrow 0$ and $a\rightarrow +\infty$ but they differ
in between. 

\subsection{Particular values}

We now provide the values of the statefinders at particular points of interest
in the Logotropic model.

(i) The pressure of the Logotropic Universe vanishes ($w=0$) at
\begin{equation}
a_w=e^{-\frac{1+B}{3B}}.
\label{stl15}
\end{equation}
At that point
\begin{equation}
q_w=\frac{1}{2},
\label{stl16}
\end{equation}
\begin{equation}
r_w=1+\frac{9B\Omega_{\rm de0}}{2\lbrack
\Omega_{\rm m0}e^{(1+B)/B}-B\Omega_{\rm de0}\rbrack},
\label{stl17}
\end{equation}
\begin{equation}
s_w=\infty.
\label{stl18}
\end{equation}
Numerically
\begin{equation}
a_w=7.00\times 10^{-42},\qquad \epsilon_w/\epsilon_0=7.97\times 10^{122},\qquad
H_0t_w=2.36\times 10^{-62},
\label{stl19}
\end{equation}
\begin{equation}
q_w=0.5,\qquad r_w=1.00, \qquad
s_w=\infty.
\label{stl20}
\end{equation}
We note that the parameter $s$ diverges at $a_w=7.00\times 10^{-42}$ while its
value in the
$\Lambda$CDM model is always $s=0$. However, this is essentially a mathematical
curiosity since the Logotropic model (which is a unification of DM and DE) may
not be justified at such small scale factors (see footnote 11).

(ii) The Logotropic Universe accelerates ($q\ge 0$) at the point $a_c$ defined
implicitly by the relation
\begin{equation}
B=\frac{\frac{\Omega_{\rm m0}}{\Omega_{\rm de0}a_c^3}-2}{3(1+2\ln a_c)}.
\label{stl21}
\end{equation}
The function $a_c(B)$ is studied in \cite{delong}. At that point
\begin{equation}
q_c=0,
\label{stl22}
\end{equation}
\begin{equation}
r_c=1+\frac{9
B\Omega_{\rm de0}}{2\lbrack \frac{\Omega_{\rm m0}}{a_c^3}+\Omega_{\rm de0}(1+3B\ln
a_c)\rbrack},
\label{stl23}
\end{equation}
\begin{equation}
s_c=-\frac{2}{3}(r_c-1)=-\frac{B}{B+1+3B\ln a_c}.
\label{stl24}
\end{equation}
Numerically 
\begin{equation}
a_c=0.574,\qquad \epsilon_c/\epsilon_0=2.17,\qquad
H_0t_c=0.515, 
\label{stl25}
\end{equation}
\begin{equation}
q_c=0,\qquad r_c=1.005, \qquad
s_c=-0.00354.
\label{stl26}
\end{equation}
For the $\Lambda$CDM model ($B=0$), we have
\begin{equation}
a_c=\left (\frac{\Omega_{\rm m0}}{2\Omega_{\rm de0}}\right )^{1/3}.
\label{stl27}
\end{equation}
Numerically
\begin{equation}
a_c=0.574,\qquad \epsilon_c/\epsilon_0=2.18,\qquad
H_0t_c=0.515,
\label{stl28}
\end{equation}
\begin{equation}
q_c=0,\qquad r_c=1, \qquad
s_c=0.
\label{stl29}
\end{equation}

(iii) The current values of the statefinders  ($a=1$) in the Logotropic
model are
\begin{equation}
q_0=\frac{1}{2}-\frac{3\Omega_{\rm de0}}{2}({B+1}),
\label{stl30}
\end{equation}
\begin{equation}
r_0=1+\frac{9
B\Omega_{\rm de0}}{2},
\label{stl31}
\end{equation}
\begin{equation}
s_0=-\frac{B}{B+1}.
\label{stl32}
\end{equation}
We emphasize that $s_0$ depends only on $B$. Therefore, the present value
of $s$ unequivocally determines $B$ independently of the values of $\Omega_{\rm m0}$
and $H_0$. Numerically
\begin{equation}
a_0=1,\qquad \epsilon_c/\epsilon_0=1,\qquad
H_0t_c=0.989,
\label{stl33}
\end{equation}
\begin{equation}
q_0=-0.593,\qquad r_0=1.01, \qquad
s_0=-0.00352.
\label{stl34}
\end{equation}
For the $\Lambda$CDM model ($B=0$), we have
\begin{equation}
a_0=1,\qquad \epsilon_c/\epsilon_0=1,\qquad
H_0t_c=0.989,
\label{stl35}
\end{equation}
\begin{equation}
q_0=-0.589,\qquad r_0=1, \qquad
s_0=0.
\label{stl36}
\end{equation}
In the $\Lambda$CDM model, $s=0$ exactly while $s_0=-0.00352$ in the
Logotropic model. Therefore, the observation of a  small negative value of $s$
would be in favor of the Logotropic model. Since $B>0$, we predict that the
distribution of measured values of $s$ about $s=0$ should be disymmetric and
should favor negatives values of $s$ with respect to positive ones. However, it
is not clear if this slight asymmetry can be observed with current precision
of measurements.

(iv) The Logotropic Universe becomes phantom ($w=-1$) at
\begin{equation}
a_M=\left (\frac{\Omega_{\rm m0}}{B\Omega_{\rm de0}}\right )^{1/3}.
\label{stl37}
\end{equation}
At that point
\begin{equation}
q_M=-1,
\label{stl38}
\end{equation}
\begin{equation}
r_M=1+\frac{9B}{2\lbrack B+1+B\ln(\Omega_{\rm m0}/B\Omega_{\rm de0})},
\label{stl39}
\end{equation}
\begin{equation}
s_M=-\frac{2}{9}(r_M-1)=-\frac{B}{B+1+B\ln(\Omega_{\rm m0}/B\Omega_{\rm de0})}.
\label{stl40}
\end{equation}
Numerically
\begin{equation}
a_M=4.75,\qquad \epsilon_M/\epsilon_0=0.7405,\qquad
H_0t_M=2.745,
\label{stl41}
\end{equation}
\begin{equation}
q_M=-1,\qquad r_M=1.015, \qquad s_M=-0.00346.
\label{stl42}
\end{equation}

\subsection{The functions $q(a)$, $r(a)$ and $s(a)$}

\begin{center}
\begin{figure}[htb]\centering
\includegraphics[width=7.6cm]{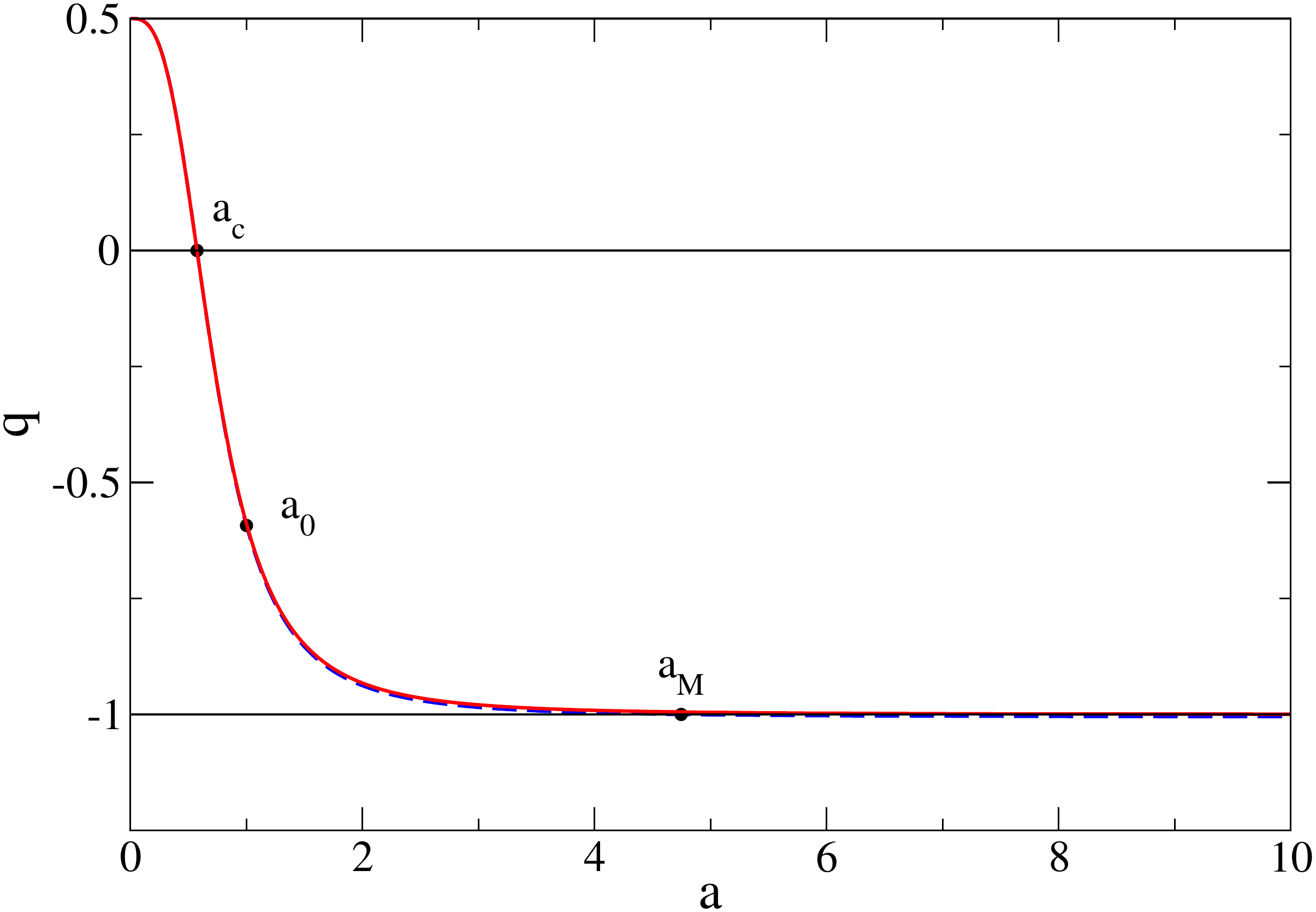}
\includegraphics[width=7.6cm]{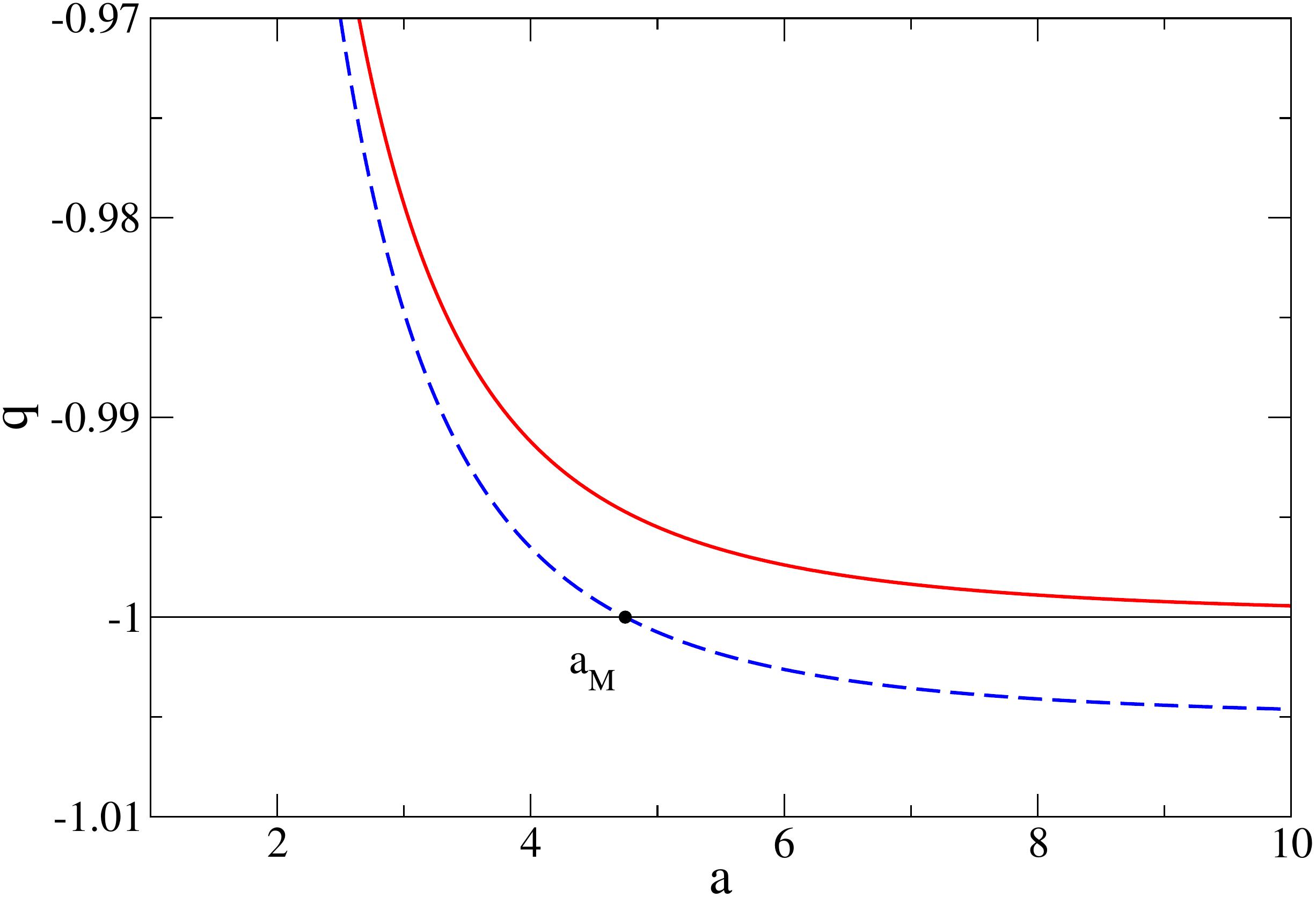}
\caption{The deceleration parameter $q$ as a function of the scale
factor for the Logotropic model (blue) and for the
$\Lambda$CDM model (red). The right panel is a zoom close to the
point where the Logotropic Universe becomes phantom.}\label{aq}
\end{figure}
\end{center}

\begin{center}
\begin{figure}[htb]\centering
\includegraphics[width=7.6cm]{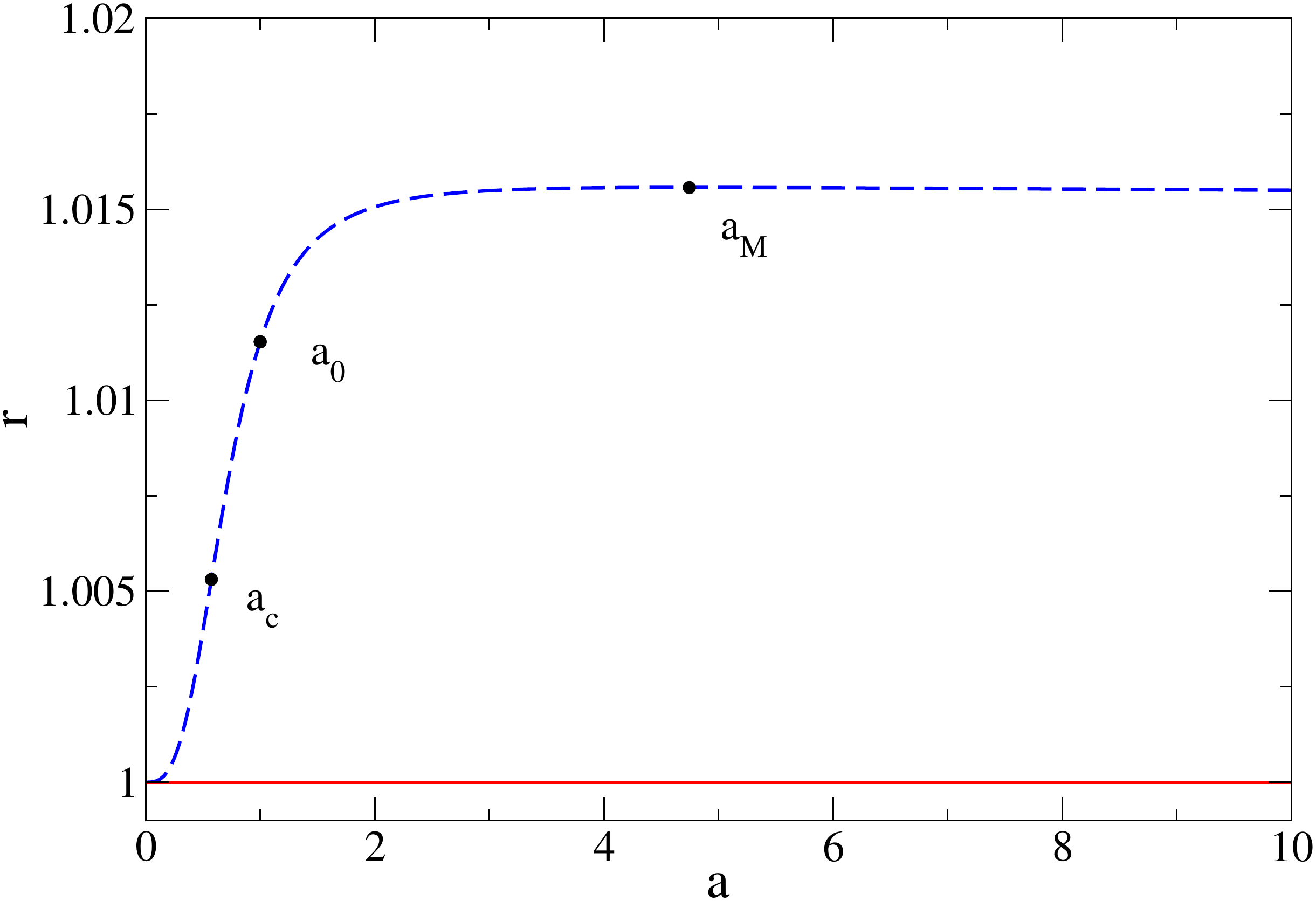}
\includegraphics[width=7.6cm]{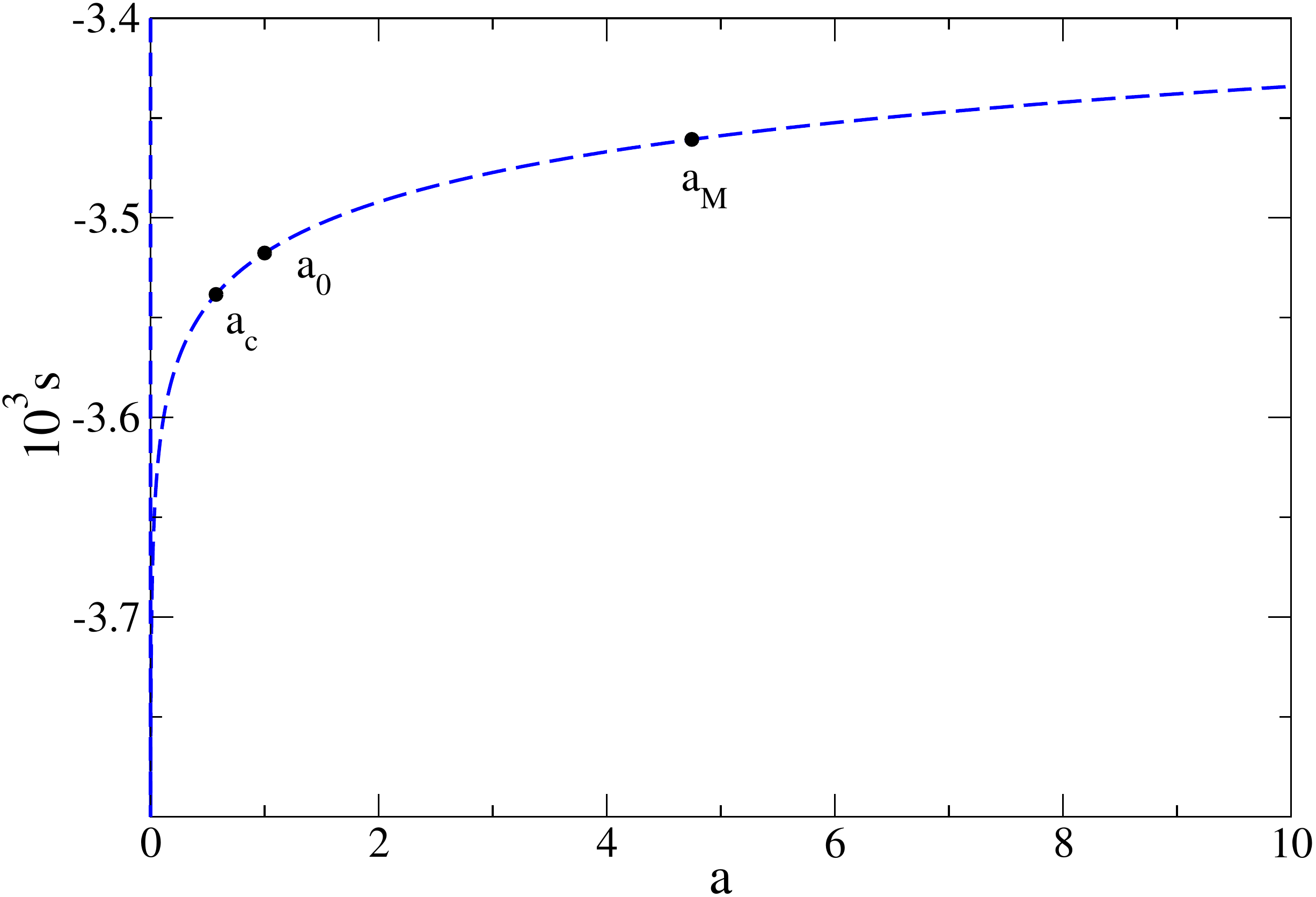}
\caption{Left: Parameter $r$ as a function of the scale
factor for the Logotropic model (blue) and for the
$\Lambda$CDM model (red). Right: Parameter $s$
(multiplied by $1000$) as a function of the scale
factor for the Logotropic model. For the $\Lambda$CDM model, $s=0$.}\label{ar}
\end{figure}
\end{center}

The differences between the  Logotropic model and the $\Lambda$CDM model are
apparent on Figs. \ref{aq} and \ref{ar} where we plot individually $q$, $r$ and
$s$ as a function of the scale factor $a$.

The function $q(a)$ (see Fig. \ref{aq}) has been studied in detail in
\cite{delong} so we remain brief. This function starts from $q_i=1/2$ at
$a_i=0$, increases,  reaches
a maximum $q_{\rm max}=0.5+1.77\times 10^{-126}$ at $a'\simeq
e^{-(2B+1)/3B}=5.02\times 10^{-42}$, decreases, takes the value $q=1/2$ at
$a_w=7.00\times 10^{-42}$ (at that point
the pressure vanishes), takes the value $q=0$ at $a_c=0.574$ (at
that point the Universe starts accelerating), takes the value $q=-1$ at
$a_M=4.75$
(at that point the Universe becomes phantom), reaches a minimum
$q_{\rm min}=-1.005$ at $a''=31.6$ (approximately $a''\simeq
[(2B+1)\Omega_{\rm m0}/B^2\Omega_{\rm de0}]^{1/3}$), increases and tends
slowly 
(logarithmically) towards $-1^{-}$ for $a\rightarrow
+\infty$. By comparison, for the $\Lambda$CDM model, the function   $q(a)$
starts from $q_i=1/2$ at $a_i=0$, decreases monotonically, takes the value $q=0$
at
$a_c=0.574$, and tends towards $-1^{+}$ for $a\rightarrow +\infty$. The
evolution of the equation of state parameter $w(a)$ can be obtained
straightforwardly from the evolution of $q(a)$ by using the relation of Eq.
(\ref{df5}).

The function $r(a)$  (see Fig. \ref{ar}, left panel) starts from $r_i=1$ at
$a_i=0$, increases,
reaches a maximum $r_M=1.015$ at $a_M=4.75$ (at that point the Universe
becomes phantom)\footnote{The parameter
$r$ can be rewritten as $r=1+9B\Omega_{\rm de0}H_0^2/2H^2$. The maximum of $r(a)$
corresponds to the minimum of $H(a)$, hence to the minimum of $\epsilon(a)$,
that is to say when
the Logotropic Universe becomes phantom.} and decreases slowly (logarithmically)
towards $1^+$ for $a\rightarrow +\infty$.

The function $s(a)$ (see Fig. \ref{ar}, right panel) starts from $s_i=0$ at
$a_i=0$, increases, tends
towards $+\infty$ as $a\rightarrow a_w^{-}=7.00\times 10^{-42}$, tends   towards
$-\infty$ as $a\rightarrow a_w^{+}=7.00\times 10^{-42}$, increases and tends
slowly  (logarithmically) towards
$0^{-}$ for $a\rightarrow
+\infty$. Since the singularity at  $a=a_w$ occurs in the very early Universe
where the Logotropic model may not be valid, we have not
represented it on the
figure. We note that for typical values of $a$, the parameter $s$ has an
approximately constant value $\sim -3.53\times 10^{-3}$, due to its slow
(logarithmic) dependence on the scale factor, which corresponds to
the value of the dimensionless Logotropic temperature $B$ (with the opposite
sign).

\subsection{The $qr$ and $sr$ planes}

\begin{center}
\begin{figure}[htb]\centering
\includegraphics[width=7.6cm]{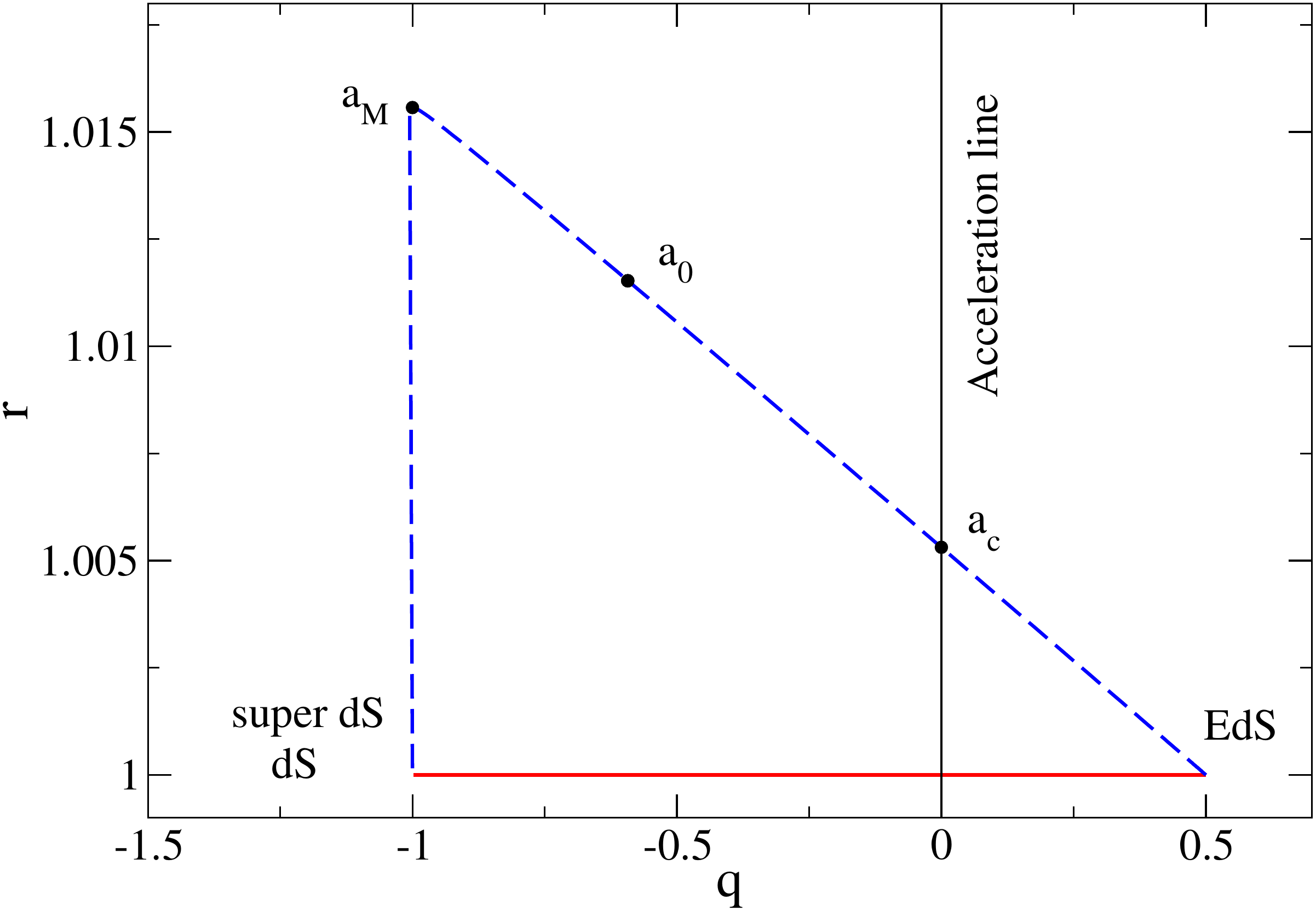}
\includegraphics[width=7.6cm]{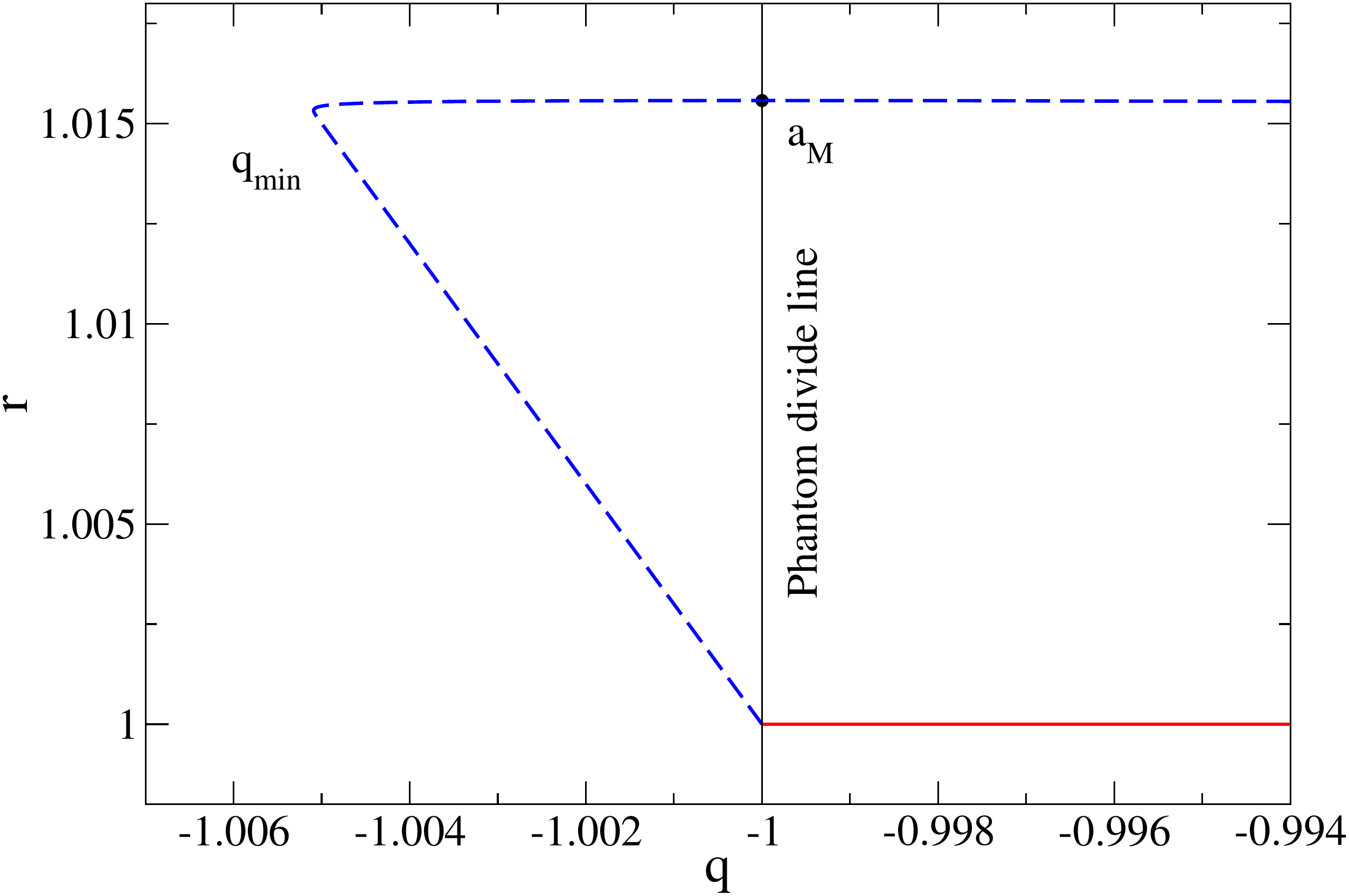}
\caption{The $qr$ trajectory for the Logotropic model (blue) and for
the
$\Lambda$CDM model (red). The right panel is a zoom  close to the
phantom divide line. We have indicated the acceleration
line where $q=0$ and the phantom divide line where $q=-1$. We have also
indicated the point where $r$ is maximum and the point where $q$ is minimum. 
}\label{qr}
\end{figure}
\end{center}

We plot the evolution trajectories of the Logotropic and $\Lambda$CDM models in
the $qr$ plane in Fig. \ref{qr}. The Logotropic model
and the $\Lambda$CDM model have different 
trajectories but evolve from a matter dominated phase (EdS) corresponding to
the point $(1/2,1)$ in the $qr$ plane to the de Sitter phase
(for the $\Lambda$CDM
model) or to the super de Sitter phase (for the Logotropic model) corresponding
to the point
$(-1,1)$ in the $qr$ plane. In this representation, the
$\Lambda$CDM model forms a segment while the evolution of the Logotropic model
is
more complex. We can discriminate
the Logotropic
model from the $\Lambda$CDM model by
observing that the dashed line (Logotropic model) runs above the
solid line ($\Lambda$CDM model) in the $qr$ plane. On the other hand, as
revealed by the zoom of Fig. \ref{qr}-b, the dashed line (Logotropic model)
crosses the phantom divide line $q=-1$, contrary to the solid line
($\Lambda$CDM model).

\begin{center}
\begin{figure}[htb]\centering
\includegraphics[width=8cm]{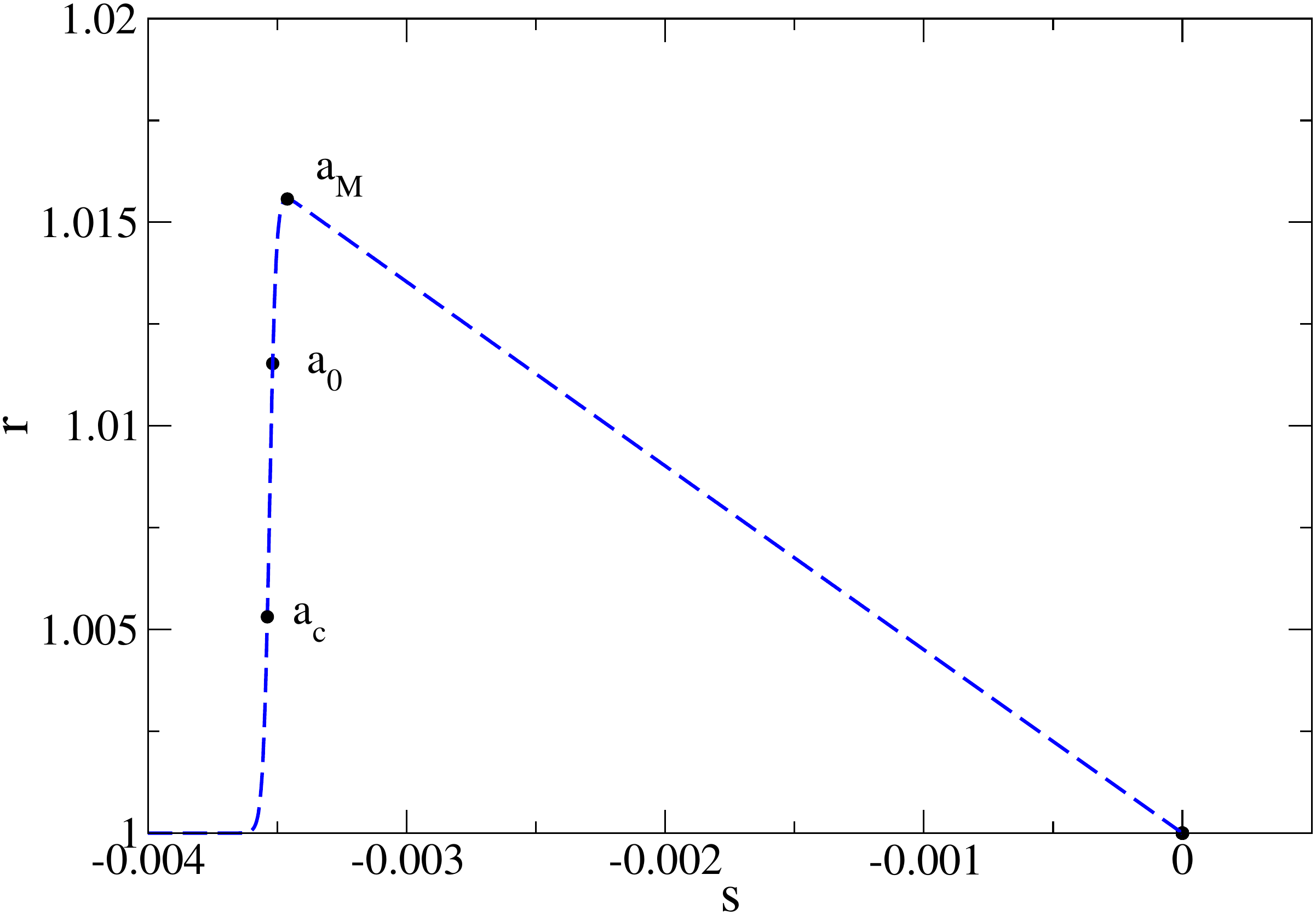}
\caption{The $sr$ trajectory for the Logotropic model. For the $\Lambda$CDM
model, $(s,r)=(0,1)$.}\label{sr}
\end{figure}
\end{center}

We have also represented the $sr$ plane in Fig. \ref{sr}. In this
representation, the $\Lambda$CDM model reduces to a point $(0,1)$ while the
Logotropic model has a more complicated evolution around that point. As
explained previously, the departure of the current value of $s$ from the
$\Lambda$CDM value $s=0$
is a direct measure of the dimensionless Logotropic temperature $B$ since
$s_0=-B/(B+1)$.

Despite minute differences in their evolution trajectories, the 
Logotropic model and the $\Lambda$CDM model are extremely close to each other so
they could be distinguished from observations only if the cosmological
parameters are calculated with a high precision of the percent
level. This precision is not reached by present-day observations. However,
even if the Logotropic model and the $\Lambda$CDM model are extremely close to
each other at the cosmological scale, they behave very differently at small
scales as discussed in the Introduction.

\section{Fine comparison between the Logotropic and $\Lambda$CDM
models}
\label{sec_fine}

In this section, we present observational constraints 
on the Logotropic model using the latest observational data from {\it Planck}
2015+Lensing+BAO+JLA+HST (see \cite{planck2015} for details of the data sets) and
compare them with the $\Lambda$CDM model. To that purpose, we consider DM and
DE as two separate/non-interacting fluids as in Sec.
\ref{sec_tf},\footnote{This simplifying assumption only affects the results
of the perturbation analysis developed in Sec. \ref{sec:PEs}. As explained in
Sec. \ref{sec_tf}, we expect to observe differences between the one fluid model
and the two fluids model at the level of the perturbations (but not at the
level of the background). The perturbation analysis for the single LDF model
will be considered in another paper.} and use the value of $B$ given in Eq.
\eqref{lt7}. We also assume
that the Universe is flat in agreement with the observations of the CMB
\cite{planck2013,planck2015}.

\subsection{Background and perturbation equations} \label{sec:PEs}

For the purpose of observational constraints, we write the expansion history of
the Logotropic model as
\begin{equation}\label{eq1}
H=H_{0}\sqrt{\Omega_{\rm r0}(1+z)^{4}+\Omega_{\rm b0}(1+z)^{3}+\Omega_{\rm dm0}(1+z)^{3}+
\Omega_{\rm de0}(1-3B\ln (1+z))},
\end{equation}
where $\Omega_{\rm r0}$, $\Omega_{\rm b0}$, $\Omega_{\rm dm0}$ and  $\Omega_{\rm de0}$ are the
present-day values of density parameters of radiation, baryonic matter, DM and
DE respectively with $\Omega_{\rm m0}=\Omega_{\rm b0}+\Omega_{\rm dm0}$ and
$\Omega_{\rm r0}+\Omega_{\rm b0}+\Omega_{\rm dm0}+\Omega_{\rm de0}=1$.
Furthermore, $z=a^{-1}-1$ is
the redshift.
We use the following perturbation equations for the density contrast and
velocity divergence in the synchronous gauge:
 \begin{eqnarray}
 \dot{\delta}_{i}&=&-(1+w_{i})(\theta_{i}+\frac{\dot{h}}{2})+\frac{\dot{w}_{i}}{1+w_{i}}\delta_{i}-3\mathcal{H}(c^2_{s,\rm eff}-c^2_{s,\rm ad})\left[\delta_{i}+3\mathcal{H}(1+w_{i})\frac{\theta_{i}}{k^2}\right],\label{eq:idelta}\\
\dot{\theta}_{i}&=&-\mathcal{H}(1-3c^2_{s,\rm eff})\theta_{i}+\frac{c^2_{s,\rm eff}}{1+w_{i}}k^2\delta_{i}-k^2\sigma_{i}\label{eq:iv},
 \end{eqnarray}
following the notations of \cite{ref:MB,ref:Hu98}. The adiabatic sound speed is given by
\begin{equation}
c^2_{s,\rm ad}=\frac{\dot{p}_{i}}{\dot{\rho}_{i}}=w_{i}-\frac{\dot{w}_{i}}{3\mathcal
{H}(1+w_{i})},
\end{equation}
where $c^2_{s,\rm eff}$ is the effective sound speed in the rest frame of the $i$th
fluid. In general, $c^2_{s,\rm eff}$ is a free model parameter, which measures the
entropy perturbations through its difference to the adiabatic sound speed via
the relation $w_{i}\Gamma_{i}=(c^2_{s,\rm eff}-c^2_{s,\rm ad})\delta^{\rm rest}_{i}$. Thus,
$w_{i}\Gamma_{i}$ characterizes the entropy perturbations. Furthermore,
$\delta^{\rm rest}_{i}=\delta_{i}+3\mathcal{H}(1+w_{i})\theta_{i}/k^2$ gives a
gauge-invariant form for the entropy perturbations. With these definitions, the
microscale properties of the energy component are characterized by three
quantities, i.e., the equation of state parameters $w_{i}$, the effective sound
speed $c^2_{s,\rm eff}$ and the shear perturbation $\sigma_i$. In this work, we
assume zero shear perturbations for the DE. Finally, for the DM and DE equation
of state parameters, we take the values of $w_{\rm m}$ and $w_{\rm de}$ defined by Eq.
(\ref{ha1}).

\subsection{Observational constraints}

We use the observational data from {\it Planck} 2015+Lensing+BAO+JLA+HST  to
perform a global fitting to the model parameter space of the Logotropic and
$\Lambda$CDM models
\begin{equation}
P\equiv\{\Omega_{\rm b} h^2,\Omega_{\rm c} h^2, 100\theta_{\rm MC},\tau, n_{\rm s},\ln[10^{10}A_{\rm s}]\}\nonumber
\end{equation}
via the Markov chain Monte Carlo (MCMC) method. Here, $\Omega_{\rm b} h^2$ and
$\Omega_{\rm c} h^2$ ($\Omega_{\rm c}$ was previously denoted $\Omega_{\rm dm}$) are
respectively the baryon and cold DM densities today, $\theta_{\rm MC}$ is an
approximation to
the angular size of the sound horizon at the time of decoupling, $\tau$ is the
Thomson scattering optical depth due to reionization, $n_{\rm s}$ is the scalar
spectrum power-law index and $\log[10^{10}A_{\rm s}]$ is the log power of the
primordial curvature perturbations \cite{planck2013}. We modified the publicly
available {\bf cosmoMC} package \cite{ref:MCMC} to include the perturbations of
DE in accordance with Eqs. (\ref{eq:idelta}) and (\ref{eq:iv}). Assuming
suitable priors on various model parameters, we obtained the constraints on the
parameters of the Logotropic and $\Lambda$CDM models displayed in Table
\ref{tab:results}.

\begin{center}
\begin{table}[h]\small\centering
\caption{Constraints on the parameters of the Logotropic and $\Lambda$CDM models from {\it Planck}
2015+Lensing+BAO+JLA+HST data. The parameter $H_0$ is in the units of km
s${}^{-1}$ Mpc${}^{-1}$}\label{tab:results}
\begin{tabular}{|l|ll|ll|}
\hline 
Model $\rightarrow$ & \multicolumn{2}{|c|}{Logotropic} &  \multicolumn{2}{c|}{$\Lambda$CDM}\\\hline
Parameter & Mean value with 68\% C.L. & Bestfit value& Mean value with 68\% C.L.
& Bestfit value \\ \hline
{$\Omega_{\rm b} h^2$} & $0.02231\pm 0.00014        $ & $0.02234 $&$ 0.02232\pm 0.00014 $ & $    0.02232$
 \\
{$\Omega_{\rm c} h^2$} & $0.1186\pm 0.0010          $ & $0.1176$& $    0.1184\pm 0.0010$ & $    0.1181$
 \\
{$100\theta_{\rm MC}$} & $1.04094\pm 0.00030        $ & $ 1.04098$& $    1.04095\pm 0.00029 $ & $    1.04107$
 \\
{$\tau$} & $0.067\pm 0.012            $ & $    0.069$& $    0.068\pm 0.012$ & $    0.078$
 \\
 {${\rm{ln}}(10^{10} A_{\rm s})$} & $3.065\pm 0.023            $ & $3.069$& $    3.067\pm 0.023 $ & $    3.087$
\\
{$n_{\rm s}$} & $0.9671\pm 0.0040          $ & $    0.9705$& $    0.9676\pm 0.0039$ & $    0.9681$
 \\

\hline
$\Omega_{\rm m0}$ & $    0.3068\pm 0.0060$ & $    0.3014$& $    0.3070\pm 0.0061$ & $    0.3049$
 \\
$H_0$ & $   67.93\pm 0.45$ & $   68.30$& $   67.87\pm 0.46 $ & $   68.02$
 \\

\hline
\end{tabular}

\end{table}
\end{center}

\begin{center}
\begin{figure}[htb]\centering
\includegraphics[width=15cm]{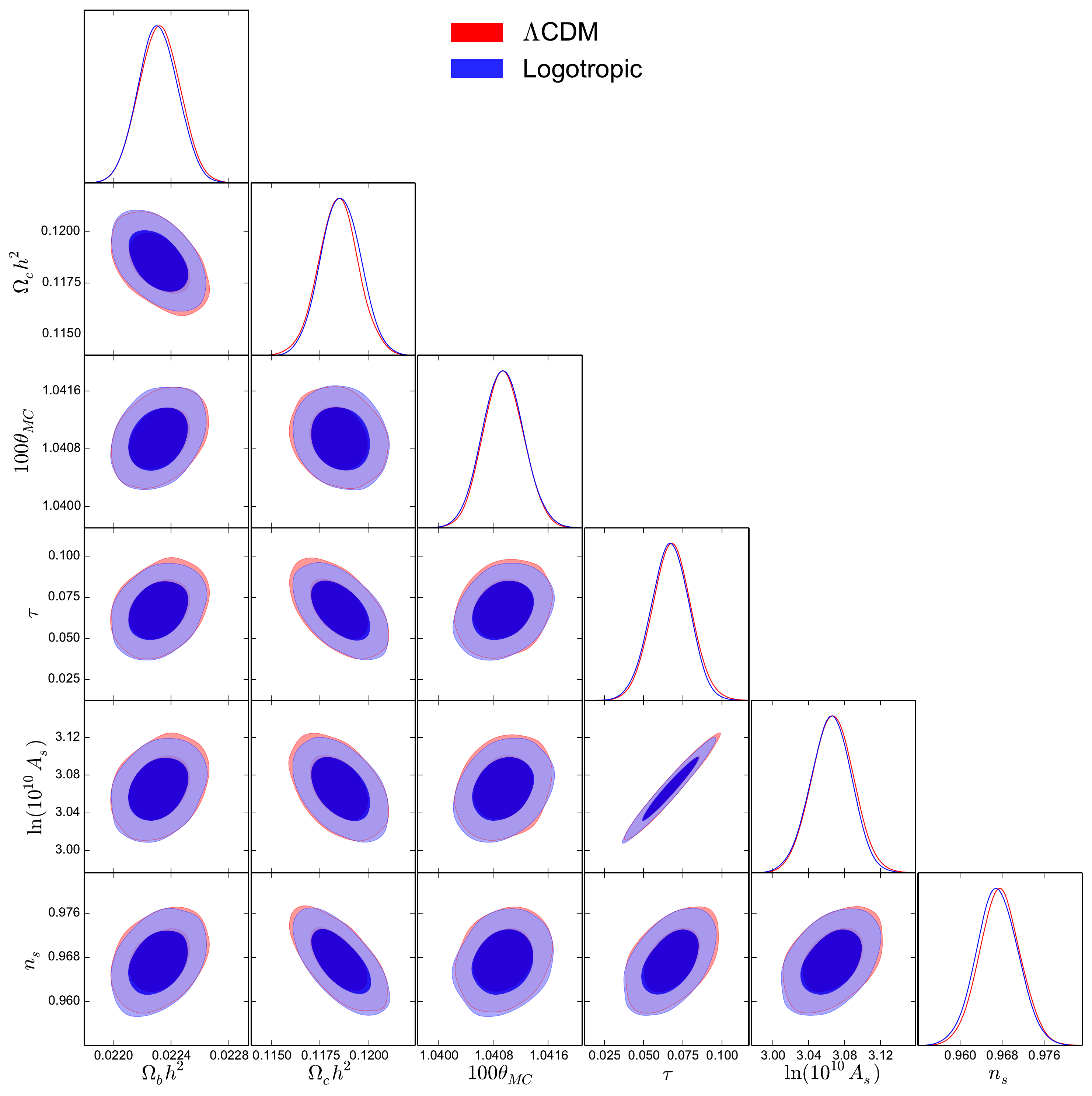}
\caption{The one-dimensional marginalized distributions of individual parameters
and two-dimensional contours  with $68\%$ C.L. and $95\%$ C.L.}\label{figcont1}
\end{figure}
\end{center}

\begin{center}
\begin{figure}[htb!]\centering
\includegraphics[width=7.6cm]{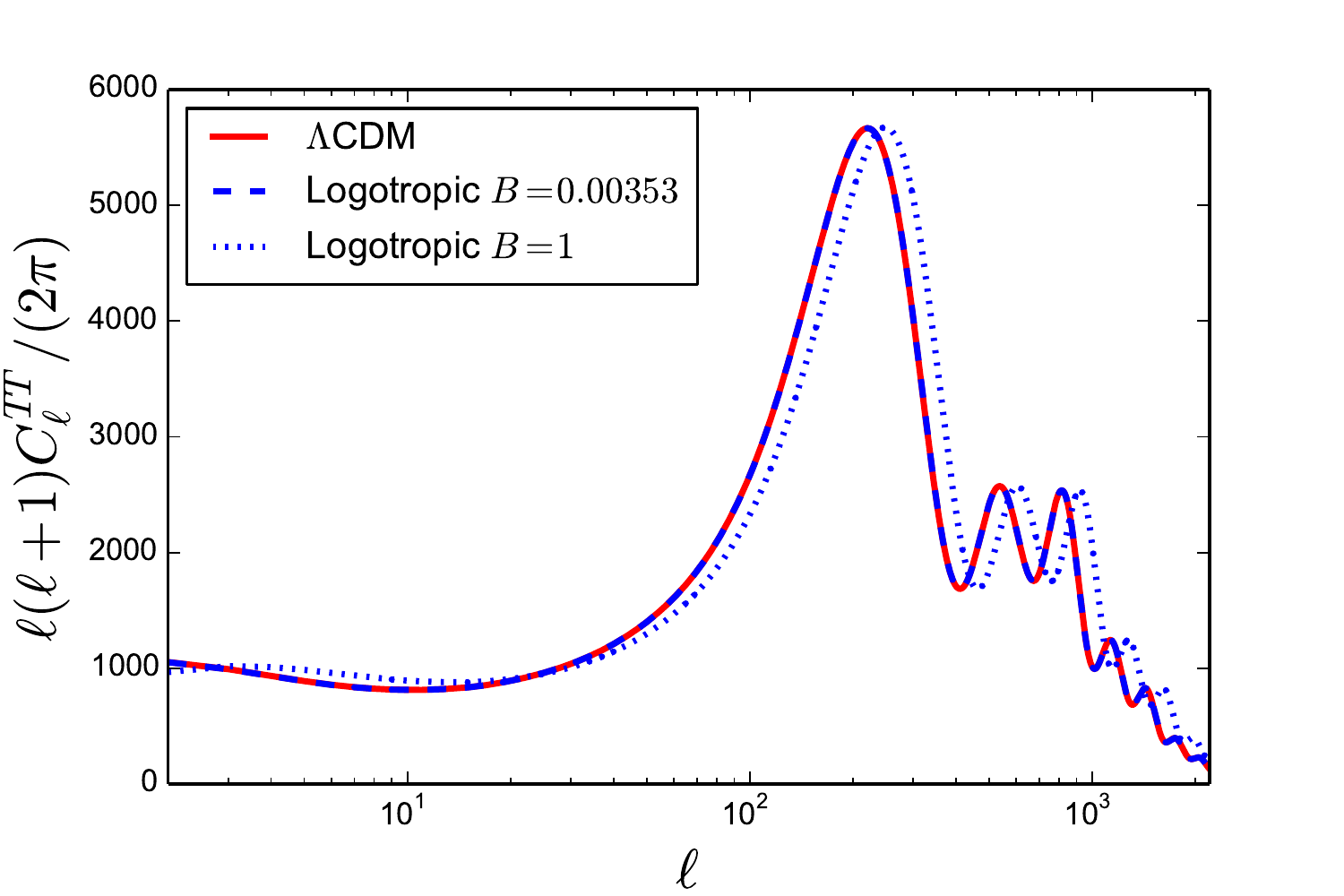}
\includegraphics[width=7.6cm]{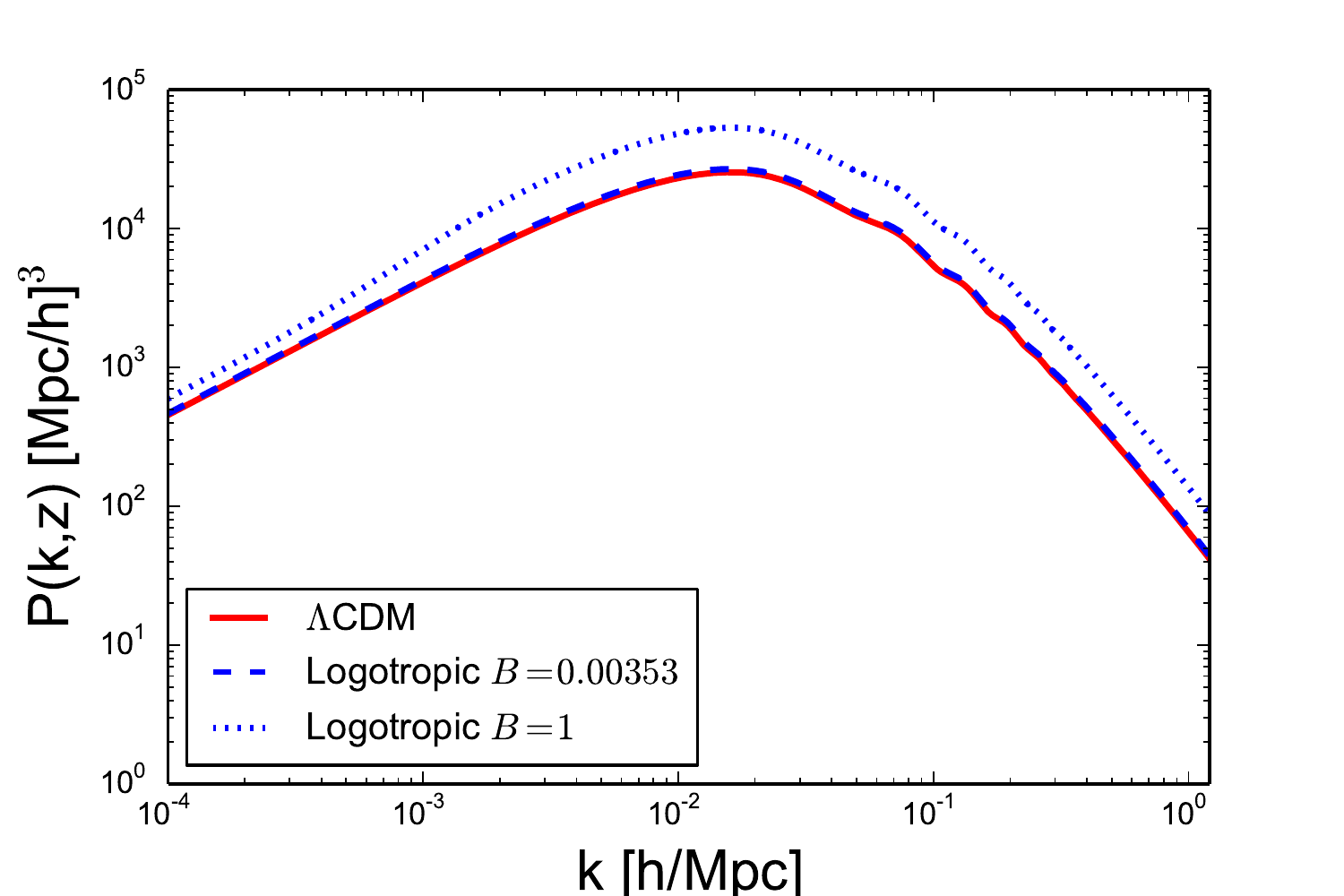}
\caption{The CMB TT power spectra (left panel) and the matter power spectra
(right panel)  at redshift $z=0$ for the $\Lambda$CDM (full red line) and
Logotropic (dashed blue line) models,
where the relevant parameters are fixed to their best fit values given in Table
\ref{tab:results}. The two curves are almost
indistinguishable, implying that the Logotropic model can
account for cosmological observations as well as the
$\Lambda$CDM model. For comparison, we have plotted the
Logotropic model with $B=1$ (blue dotted line) which presents
a strong deviation from the $\Lambda$CDM model. This confirms
that the parameter $B$ must be sufficiently small, such as the predicted value
$B=3.53\times 10^{-3}$, to account for the observations
\cite{delong,decourt}.}\label{fig:clspk}
\end{figure}
\end{center}

In Fig. \ref{figcont1}, we show one-dimensional marginalized distributions of
individual parameters and two-dimensional contours  with $68\%$ C.L. and $95\%$
C.L. for the model parameters under consideration. The CMB TT power spectra and
matter power spectra  at redshift $z=0$ for the $\Lambda$CDM and Logotropic
models are displayed in Fig. \ref{fig:clspk}, where the relevant parameters
are fixed to their best fit values as given in Table \ref{tab:results}. From
Table \ref{tab:results},  Fig. \ref{figcont1} and Fig. \ref{fig:clspk}, we
notice that there is no significant difference between the $\Lambda$CDM and
Logotropic models at the present epoch, as expected. However, the Logotropic
model will behave differently from the $\Lambda$CDM model in the
future evolution of the Universe as the logarithmic term will eventually be
significant for larger values of $a$. In the following subsection, we quantify
this difference by testing the evolutionary behavior of some parameters
pertaining to the two models under consideration.

\subsection{Statefinders and behavior of dark energy}

The statefinder analysis is done as follows. The evolution trajectories of
  the $\{q,r\}$ and $\{s,r\}$ pairs are plotted in $qr$ and $sr$ planes. Since
the jerk parameter of the $\Lambda$CDM model is $r=1$ whilst $s=0$, the
$\Lambda$CDM model is represented by the point $(0,1)$ in the $sr$ plane. On the
other hand, $q$ varies from $1/2$ to $-1$ in the $\Lambda$CDM model. Therefore,
the $qr$ trajectory for the $\Lambda$CDM model is a straight line segment going
from $(1/2,1)$ to $(-1,1)$ in the $qr$ plane. By plotting the $qr$ and $sr$
trajectories for the models under consideration, one can easily observe the
difference between their evolutionary behavior.

\begin{figure}[htb!]
\centering
\includegraphics[width=7.55cm]{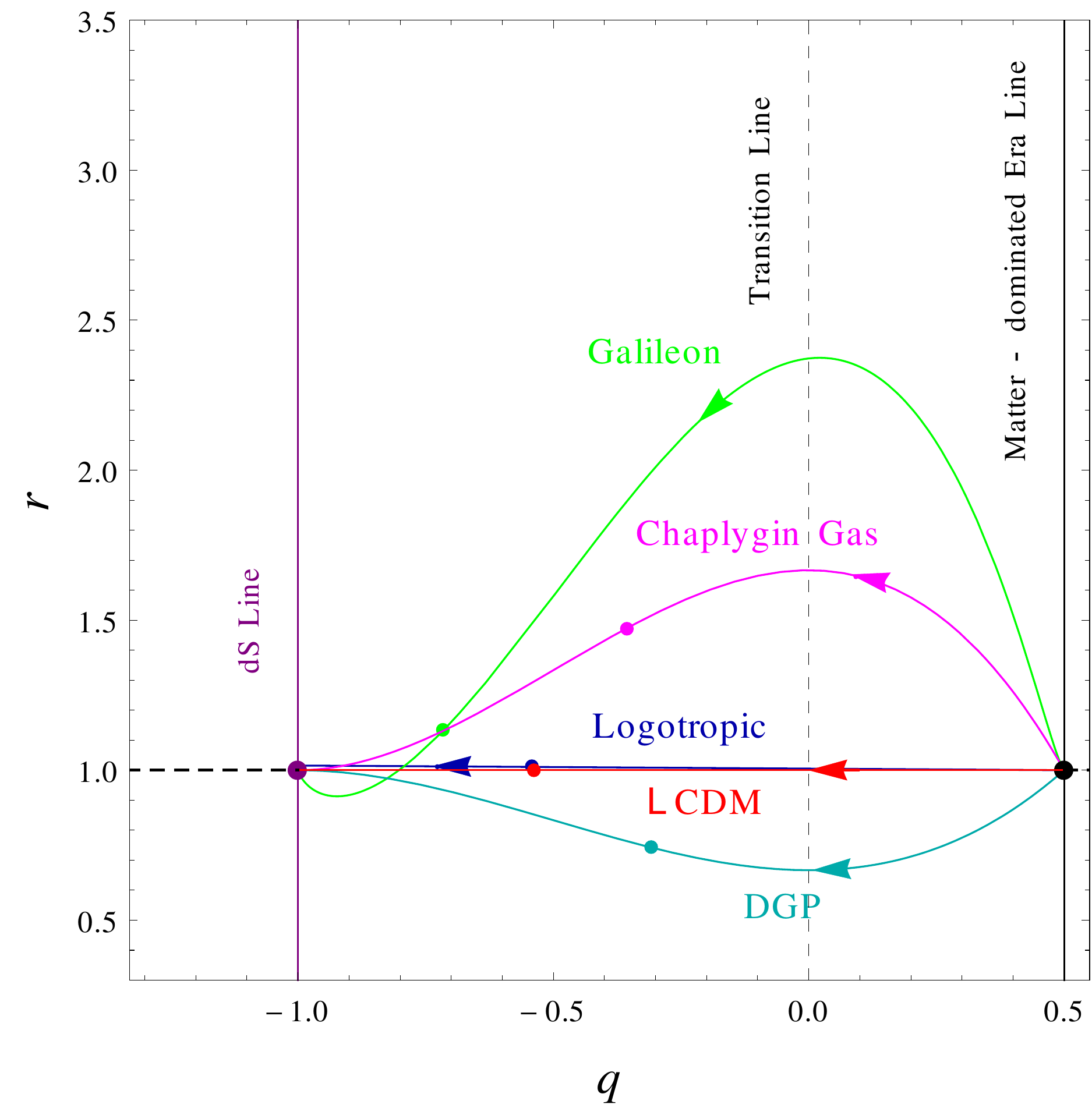}
\hspace{0.1cm}
\includegraphics[width=7.55cm]{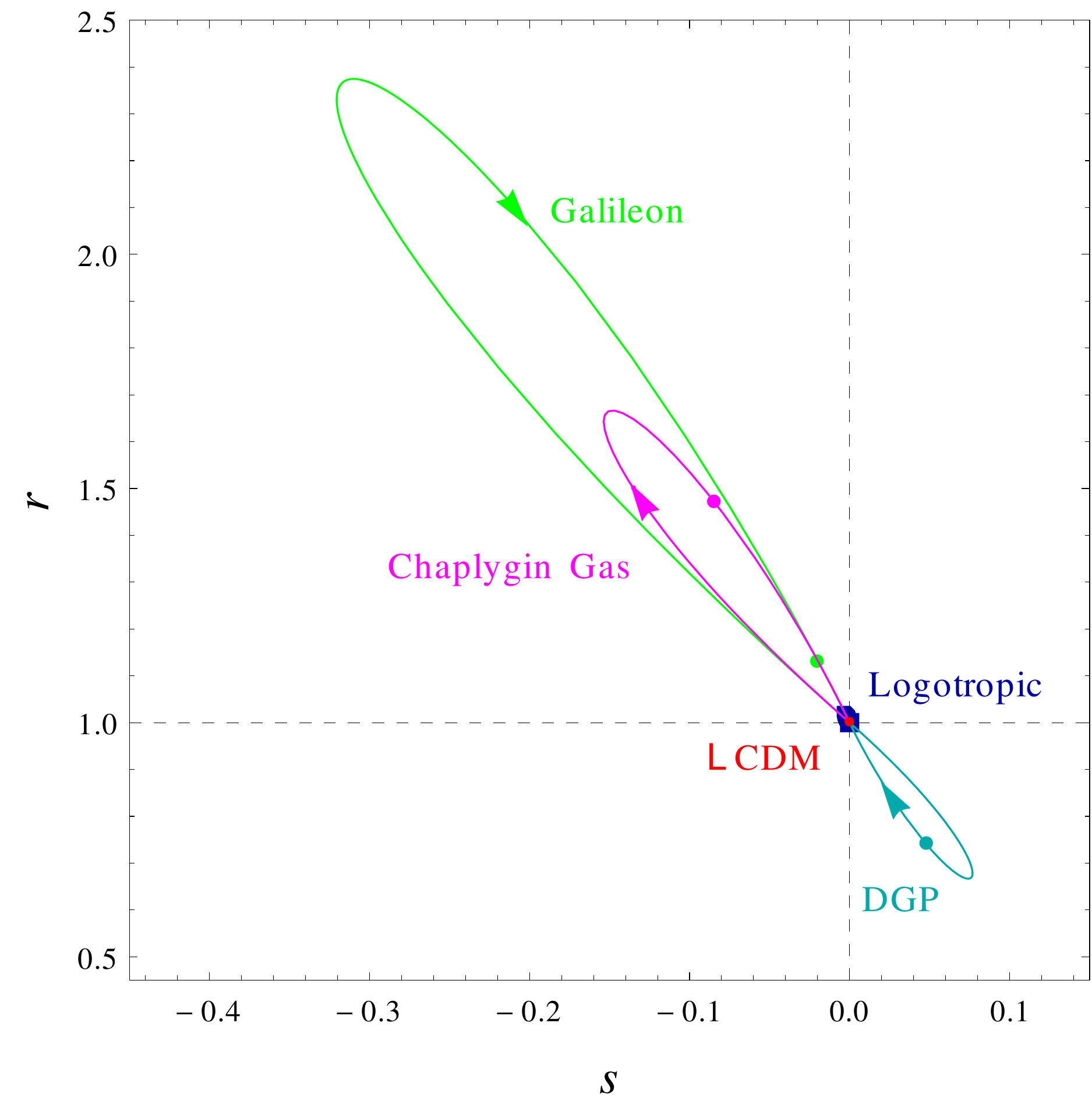}
\caption{\footnotesize{The left and right panels respectively show the evolution
of the $qr$ and $sr$ trajectories for various cosmological models. In both
panels, the blue curve corresponds to the Logotropic model.  Red, cyan, magenta
and green curves stand for $\Lambda$CDM, DGP, Chaplygin gas and Galileon models,
respectively. The direction of evolution is shown by the arrows on the curves
while the dots on the curves are used to indicate the present values of the
corresponding $\{q,r\}$ and $\{s,r\}$ pairs. All the models under consideration
evolve from the matter dominated (EdS) phase (black dot $(1/2,1)$ in the $qr$
plane)
to the de Sitter phase (purple dot $(-1,1)$ in the $qr$ plane). The Logotropic
model shows a good consistency with the $\Lambda$CDM model till
the blue dot on  the $qr$ curve. But later on, it is
visible that the blue curve departs from the red
line (see the zoom in Fig. \ref{zoomedqr}).}}
\label{fig:statefinders}
\end{figure}

\begin{center}
\begin{figure}[htb]\centering
\includegraphics[width=7cm]{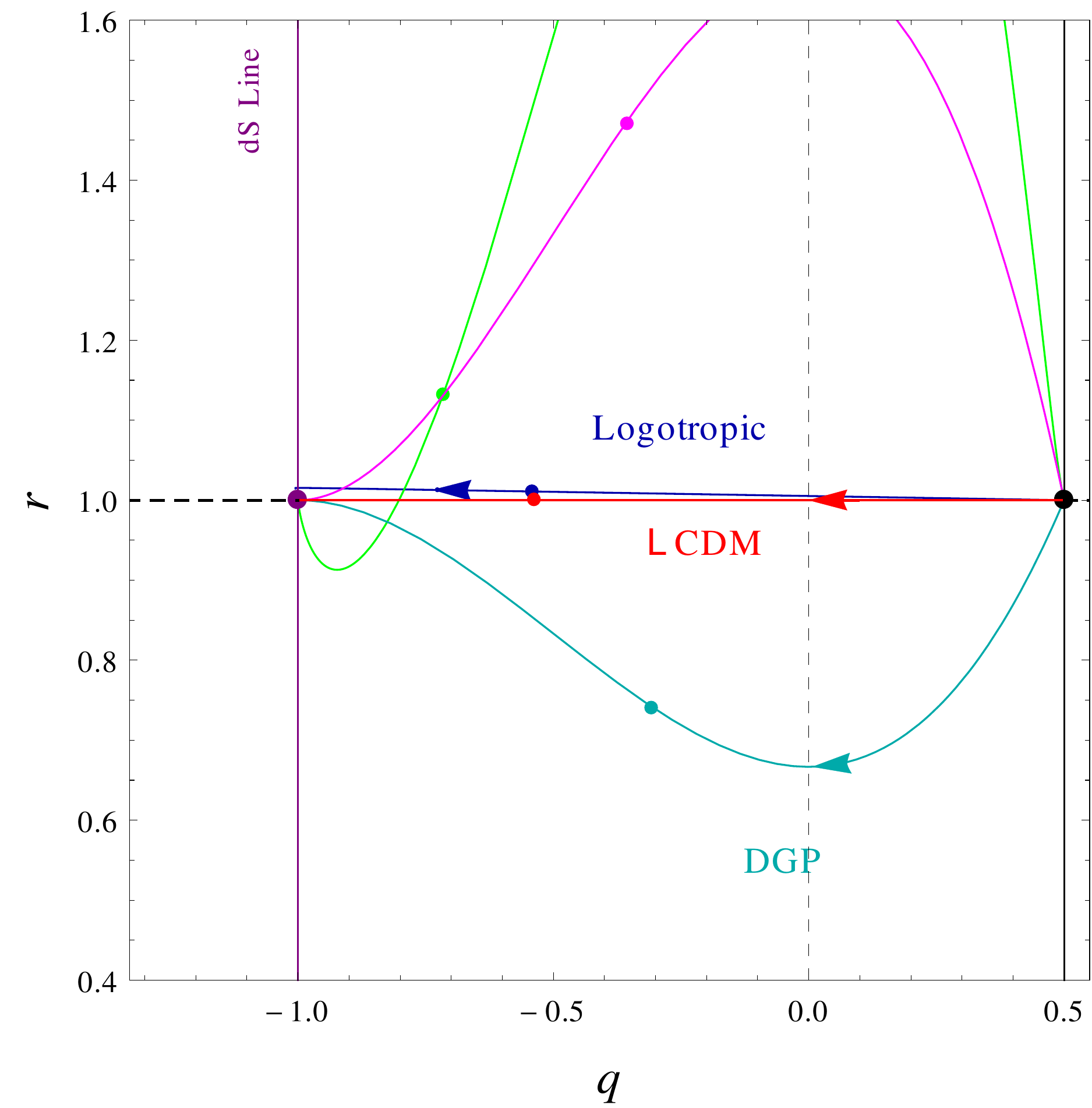}
\caption{Zoom of Fig. \ref{fig:statefinders} enlightening the
difference between the Logotropic and $\Lambda$CDM models.}\label{zoomedqr}
\end{figure}
\end{center}

We plot the evolution trajectories of the Logotropic and $\Lambda$CDM models in
the
$qr$ and $sr$ planes in the left and right panels of Fig. \ref{fig:statefinders}
by considering the best fit values of the model parameters given in Table
\ref{tab:results} from observations. For the sake of comparison, we also plot
the $qr$ and $sr$ trajectories of some popular models such as
the DGP \cite{dgp00}, Chaplygin gas \cite{chap01} and Galileon \cite{gal10}
models (see
\cite{Sami12} for the statefinder analysis of these models).

\begin{figure}[htb!]
\centering
\includegraphics[width=7.55cm]{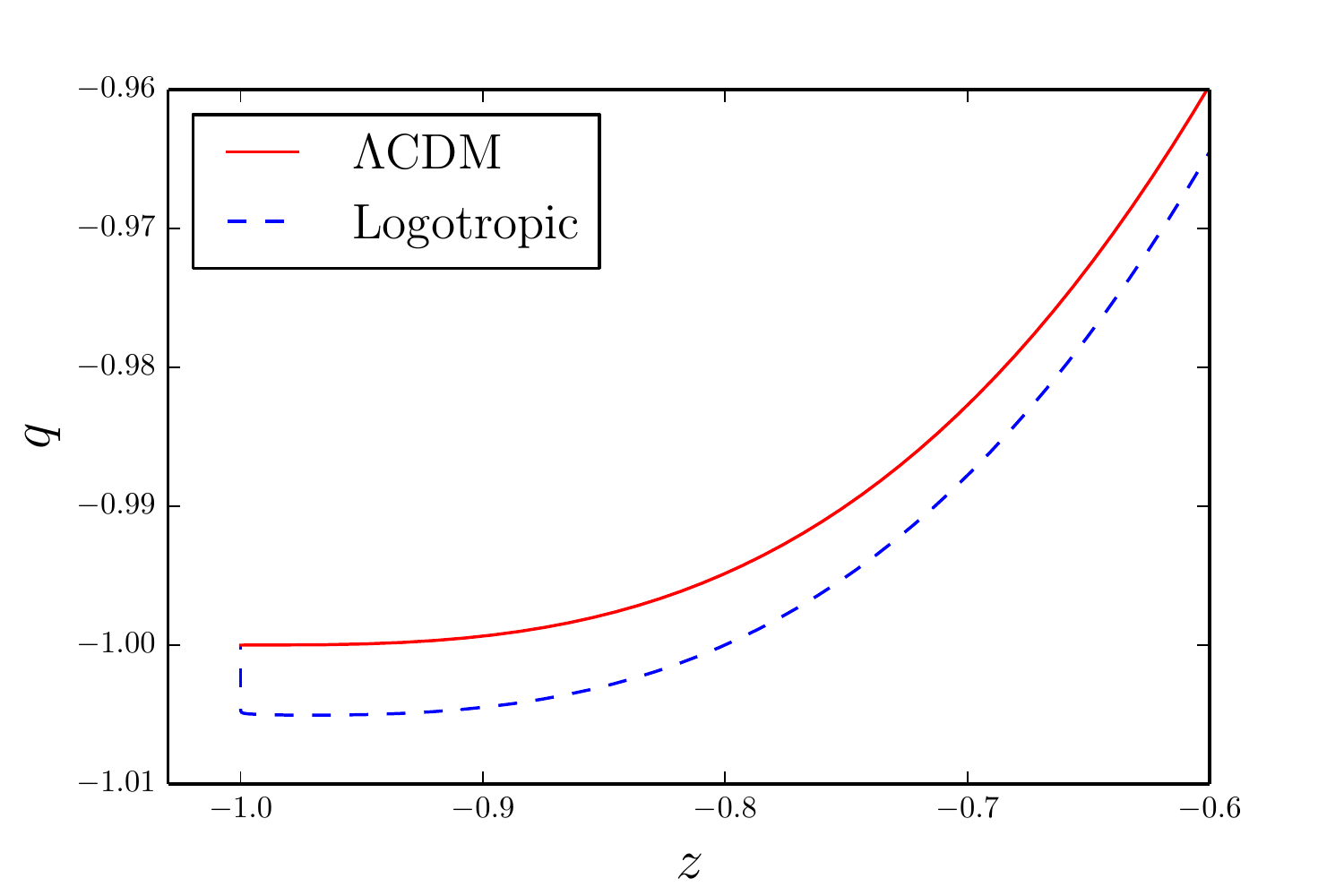}
\hspace{0.1cm}
\includegraphics[width=7.55cm]{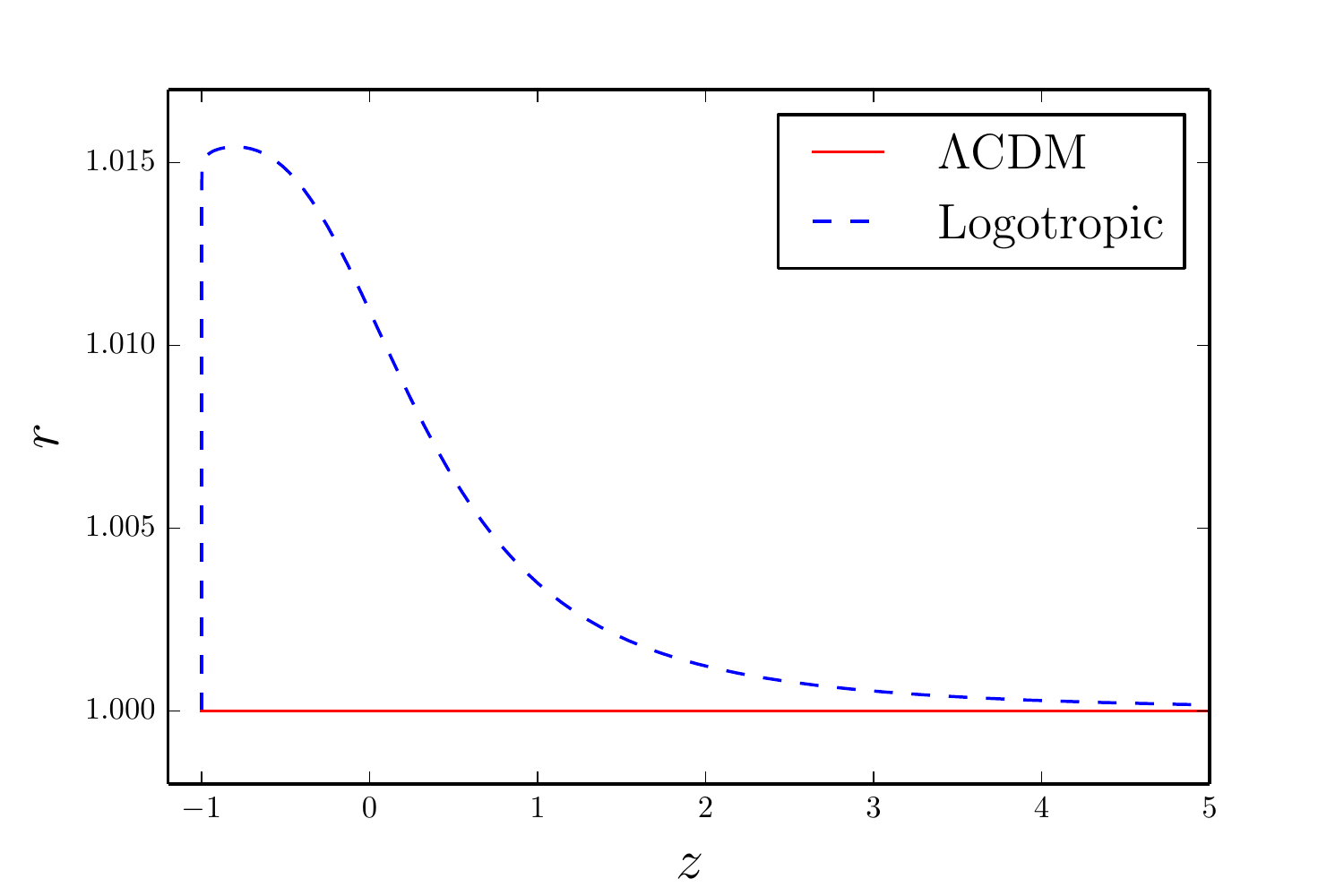}
\caption{\footnotesize{The left and right panels respectively show the variation
of $q$ and $r$ vs $z$ in the $\Lambda$CDM and Logotropic
models. }}
\label{fig:qzrz}
\end{figure}

\begin{figure}[htb!]
\centering
\includegraphics[width=7.55cm]{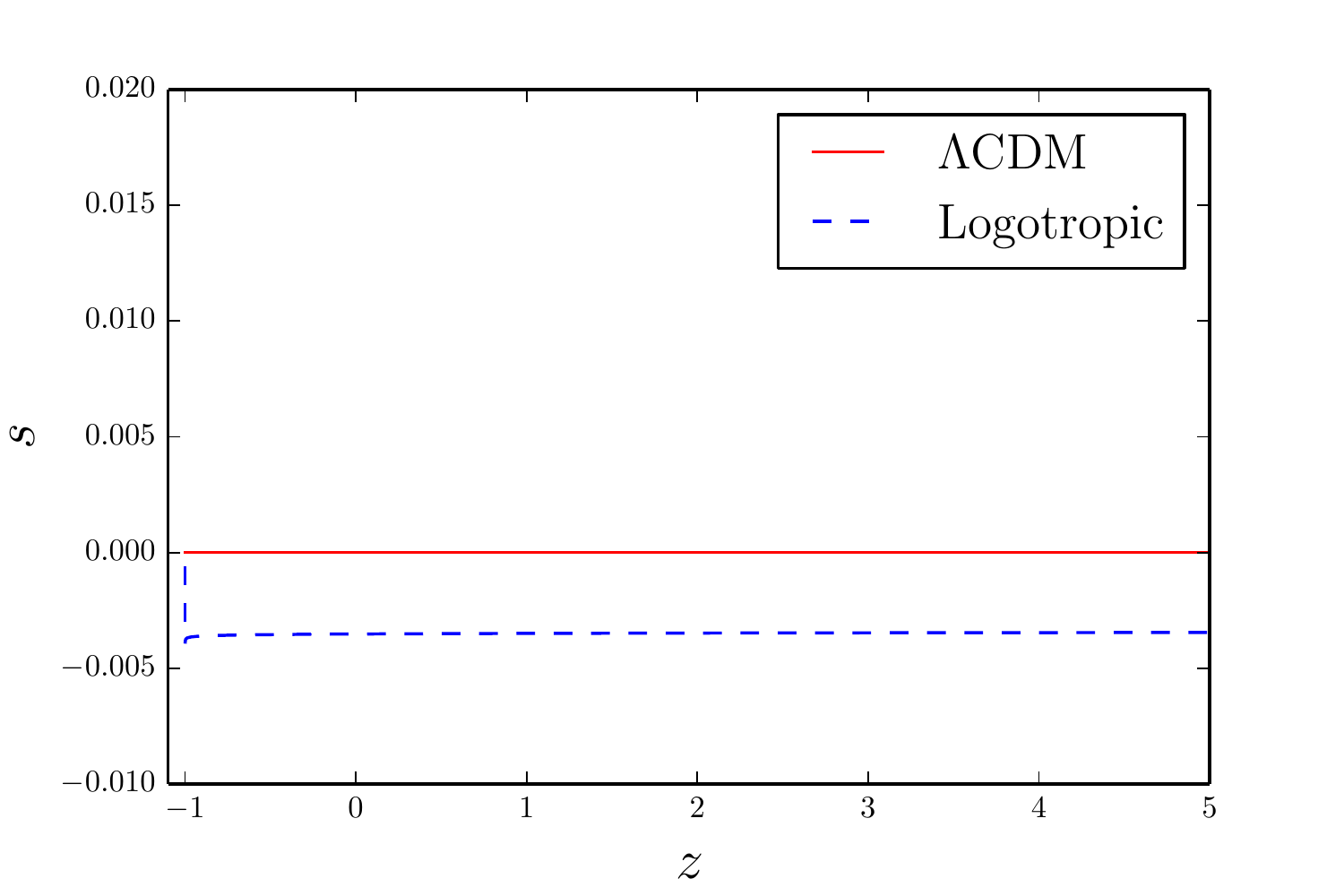}
\includegraphics[width=7.55cm]{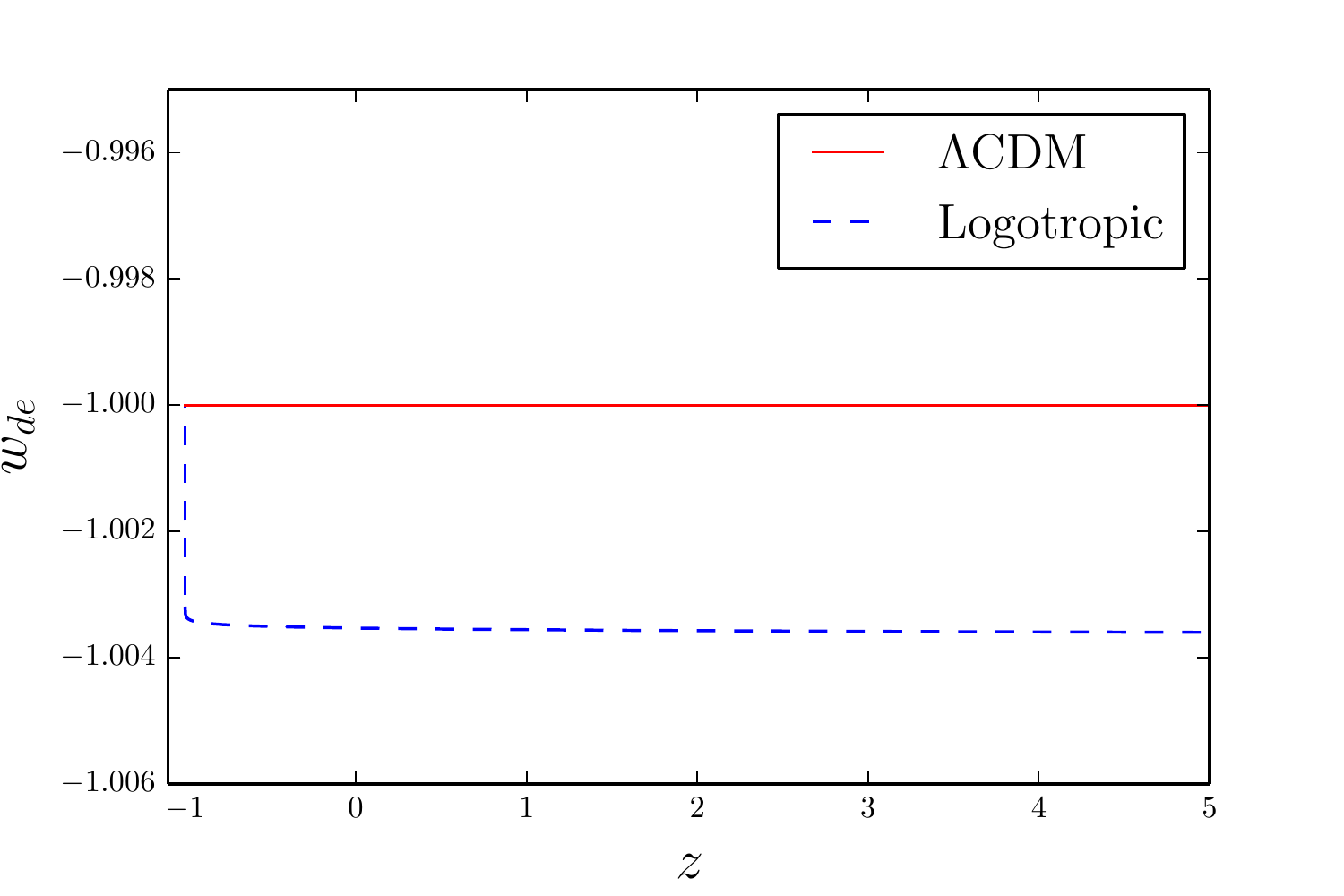}
\caption{\footnotesize{Variation
of $s$ vs $z$ in the $\Lambda$CDM and Logotropic
models (left). Variation
of $w_{\rm de}$ vs $z$ in the $\Lambda$CDM and Logotropic models (right). }}
\label{fig:sz}
\end{figure}

We see that all the models have different evolution trajectories
but evolve from the matter dominated (EdS) phase (black dot $(1/2,1)$ in the
$qr$
plane) to the de Sitter phase (purple dot $(-1,1)$ in the $qr$ plane). Of all
these models, the Logotropic model is the closest to the  $\Lambda$CDM model. We
can discriminate the Logotropic model from the $\Lambda$CDM model by
observing that the blue curve (Logotropic model) evolves differently from the
red line ($\Lambda$CDM model) in the $qr$ plane. The blue curve can be seen to
run slightly above the red line, especially after the blue dot corresponding to
the current Universe. In addition, the blue curve crosses the de Sitter line
$q=-1$ (phantom divide) before finally converging towards the de Sitter point
$(-1,1)$, as explained in Sec. \ref{sec_stl}. To display this
behavior more clearly, we plot $q$, $r$ and
$s$ separately in Figs. \ref{fig:qzrz} and \ref{fig:sz} (left) where the
departure of the Logotropic model from the $\Lambda$CDM model can clearly be
observed in the future evolution of the Universe.

We numerically find that the transition of the Universe from deceleration 
to acceleration in the $\Lambda$CDM model takes place at $z_{c}=0.667$ while in
the Logotropic model the transition redshift is 
$z_{c}=0.658$.

In Fig. \ref{fig:sz} (right), we show the evolution of the effective DE
equation of state
parameter $w_{\rm de}$ defined by Eq. (\ref{ha1}) vs $z$ in the $\Lambda$CDM
and Logotropic models. The effective DE equation of state
parameter
$w_{\rm de}$ stays less than $-1$ during the evolution of the Logotropic Universe
indicating the phantom nature of DE in the Logotropic model (its current value
is $w_{\rm de0}=-1-B=-1.00353$). In the $\Lambda$CDM model,
$w_{\rm de}=-1$.

\begin{figure}[htb!]
\centering
\includegraphics[width=7.55cm]{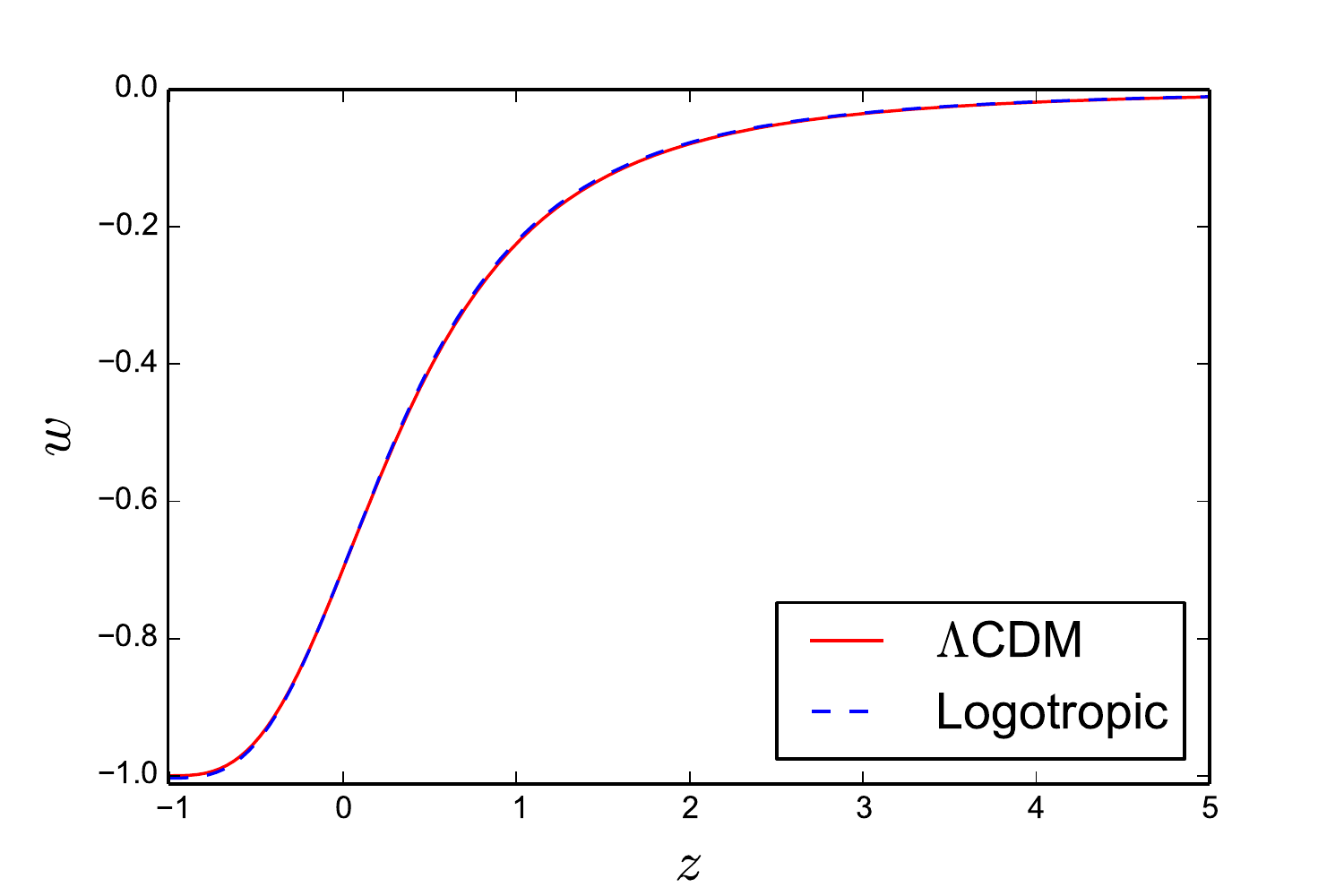}
\hspace{0.1cm}
\includegraphics[width=7.55cm]{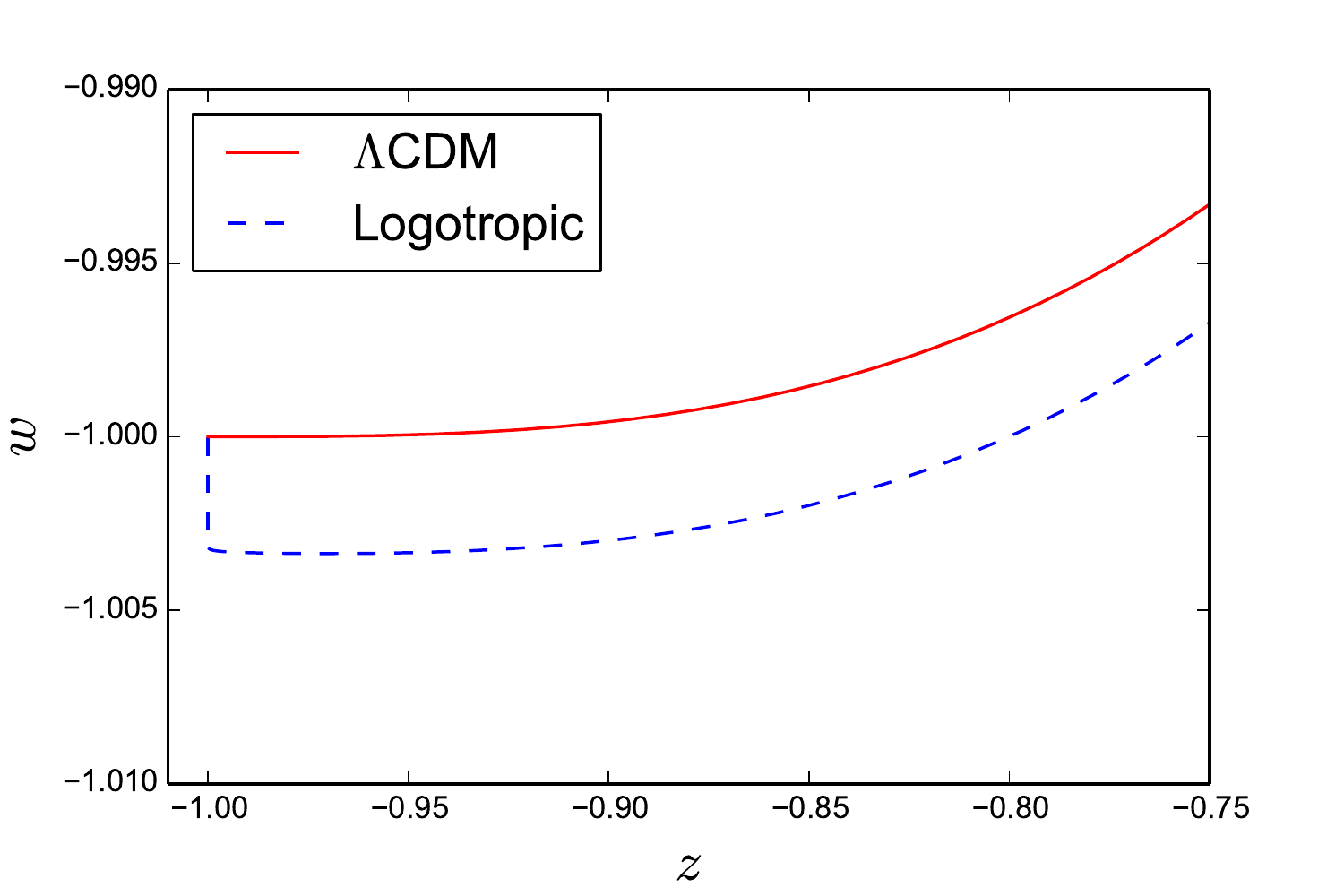}
\caption{\footnotesize{Variation
of $w$ vs $z$ in the $\Lambda$CDM and Logotropic models.}}
\label{toto}
\end{figure}

In Figs. \ref{toto}, we show the evolution of the
equation of state parameter $w$  defined by Eq. (\ref{be6w})
vs $z$ in the $\Lambda$CDM
and Logotropic models.
We note that $w$ in the Logotropic model is close to $0$ at
large redshifts, decreases (its current value is $w_0=-0.7011$), drops below
$-1$ at $z_M=-0.798$, reaches its minimum $w_{\rm min}=-1.003$ at $z''=-0.968$,
and asymptotically tends towards $-1^{-}$ as $z\rightarrow -1$. Thus, a phantom
flip
signature is observed in the future evolution of the Logotropic Universe, which
is not the case in the $\Lambda$CDM model ($w$ starts from $0$
and decreases monotonically towards $-1^+$ as $z\rightarrow -1$; its current
value is $w_0=-0.6951$).

\subsection{Numerical applications} 
\label{sec:na}

In this section, we provide the values of some quantities of cosmological
interest at different epochs in the evolution of the Universe (see Table
2). We make the numerical application for  $B=0$ corresponding
to the $\Lambda$CDM
model (see Sec. \ref{sec_lcdm}),
and for $B=3.53\times 10^{-3}$ corresponding to the Logotropic model (see
Sec. \ref{sec_lt}). We take the values of the cosmological parameters $H_0$ and
$\Omega_{\rm m0}$ obtained from observations (see  Table \ref{tab:results}).

\begin{table*}[t]
\centering
\begin{tabular}{|c|c|c|c|c|c|c|}
\hline
 & $B=0$ ($\Lambda$CDM) &  $B=3.53\times 10^{-3}$ \\
\hline
$a_w$ &  &  $7.004\times10^{-42}$ \\
\hline
$(\epsilon/\epsilon_0)_w$ &  & $8.771\times 10^{122}$  \\
\hline
$t_w$ (Gyrs) &  & $3.225\times 10^{-61}$  \\
\hline
$q_w$ &  &  $1/2$ \\
\hline
$w_w$ &  &  $0$ \\
\hline
$r_w$ &  &  $1.000$ \\
\hline
$s_w$  &  &  $\infty$ \\
\hline
$a_c$ & $0.6031$ & $0.5998$  \\
\hline
$(\epsilon/\epsilon_0)_c$ & $2.085$ & $2.092$  \\
\hline
$t_c$ (Gyrs) & $7.574$ & $7.528$  \\
\hline
$q_c$ & $0$ & $0$  \\
\hline
$w_c$ & $-1/3$ & $-1/3$  \\
\hline
$r_c$ & $1$ & $1.005$ \\
\hline
$s_c$  & $0$  & $-0.003537$  \\
\hline
$t_0$ (Gyrs) & $13.81$ & $13.80$   \\
\hline
$q_0$  & $-0.5427$ &  $-0.5516$ \\
\hline
$w_0$  & $-0.6951$ &   $-0.7011$\\
\hline
$r_0$ & $1$ & $1.011$\\
\hline
$s_0$  & $0$ & $-0.003518$  \\
\hline
$a_M$ &  &  $4.963$ \\
\hline
$(\epsilon/\epsilon_0)_M$ &  & $0.7129$  \\
\hline
$t_M$ (Gyrs) &  & $40.09$  \\
\hline
$q_M$  &  & $-1$  \\
\hline
$w_M$  &  &  $-1$ \\
\hline
$r_M$ &  & $1.016$\\
\hline
$s_M$  &  &  $-0.003459$\\
\hline
\end{tabular}
\label{tab:table2}
\caption{Numerical values of some quantities of
cosmological interest (scale factor $a$, energy density $\epsilon$, time $t$,
deceleration parameter $q$, equation of state parameter $w$, statefinders $r$
and $s$) at
different periods of the evolution of the Universe.  We recall
that $t_w$ is the time at which the Logotropic pressure becomes negative, $t_c$
is the time at which the Universe accelerates, $t_0$ is the age of the Universe,
and $t_M$ is the time at which the Logotropic Universe becomes phantom. These
results update those of Table 1 of \cite{delong}. In this Table, we have
neglected the contribution of radiation. } 
\end{table*}

For the $\Lambda$CDM model, 
$\Omega_{\rm m0}=0.3049$, $\Omega_{\rm de0}=0.6951$, 
$H_0=68.02 \, {\rm km}\,  {\rm s}^{-1}\, {\rm Mpc}^{-1}=2.204\, 10^{-18} \,
{\rm s}^{-1}$, $\epsilon_0/c^2={3H_0^2}/{8\pi
G}=8.691\times 10^{-24}\, {\rm
g}\, {\rm m}^{-3}$, $\epsilon_{\rm m0}/c^2=\Omega_{\rm m0}\epsilon_0/c^2=2.650\times
10^{-24}\,
{\rm g}\, {\rm
m}^{-3}$, and $\epsilon_{\rm de0}/c^2=\Omega_{\rm de0}\epsilon_0/c^2=6.041\times
10^{-24}\, {\rm g}\, {\rm m}^{-3}$. 

For the Logotropic model,
$\Omega_{\rm m0}=0.3014$, $\Omega_{\rm de0}=0.6986$, 
$H_0=68.30 \, {\rm km}\,  {\rm s}^{-1}\, {\rm Mpc}^{-1}=2.213\, 10^{-18} \,
{\rm s}^{-1}$, $\epsilon_0/c^2={3H_0^2}/{8\pi
G}=8.763\times 10^{-24}\, {\rm
g}\, {\rm m}^{-3}$, $\epsilon_{\rm m0}/c^2=\Omega_{\rm m0}\epsilon_0/c^2=2.641\times
10^{-24}\,
{\rm g}\, {\rm
m}^{-3}$, and $\epsilon_{\rm de0}/c^2=\Omega_{\rm de0}\epsilon_0/c^2=6.122\times
10^{-24}\, {\rm g}\, {\rm m}^{-3}$. These values improve those given in
Sec. \ref{sec_lt}. If we recompute $B$ and $A$ from Eq. (\ref{lt6})
with these more accurate values we obtain 
\begin{equation}
B=3.535\times 10^{-3},\qquad A=1.945\times 10^{-9}
\, {\rm g}\, {\rm m}^{-1}\, {\rm s}^{-2}. 
\label{lt7new}
\end{equation}
We see that the value of $B$ is not changed from the one given by Eq.
(\ref{lt7}). This shows the robustness of this value for the reason explained 
in Sec. \ref{sec_lt}. On the other hand, the value of $A$ is slightly
changed since it depends more sensibly than $B$ on the measured values of  
$H_0$ and
$\Omega_{\rm m0}$. However, we did not need the value of $A$ in the data
analysis,
so that our results are not altered.

\section{Conclusion}
\label{sec_conclusion}

In this paper, we have compared the Logotropic and $\Lambda$CDM models at large
(cosmological) scales. This comparison is interesting because
these two models are very different from each other on a theoretical point of
view. As
anticipated in \cite{delong,decourt}, the two models give results that
are very close to each other up to the current epoch. Our
detailed study shows that the difference is at the percent level (not smaller
and not larger).
This is
smaller than present-day cosmological precision. Therefore, the two models are
indistinguishable at present. Still, they will differ from
each other in the far future, in about $25$ Gyrs, since the Logotropic Universe
will
become phantom unlike the $\Lambda$CDM Universe. The closeness
of the results in the period where we can compare these two models with the 
observations implies that the Logotropic model is viable. Therefore, we
cannot reject the possibility that our Universe will become phantom in the
future. Indeed, the Logotropic model is an example of phantom Universe that is
consistent with the observations since it leads to results that are
almost indistinguishable from the $\Lambda$CDM model up to the current epoch.
This is very
different from the other models considered in Fig. \ref{fig:statefinders} which
deviate more strongly from the $\Lambda$CDM model. It may be argued that these
models are not consistent with the observations since they are ``too far''
from the standard $\Lambda$CDM model.

In a sense, it is obvious that the Logotropic model produces results that are
consistent with the observations since it depends on a parameter $B$ in such a
way that the $\Lambda$CDM model is recovered for $B=0$. Therefore, by taking $B$
sufficiently small, we are guaranteed to reproduce the results of the
$\Lambda$CDM model.\footnote{Inversely, too large values of
$B$ lead to
unacceptable deviations from the $\Lambda$CDM model as shown in Fig.
\ref{fig:clspk} for the CMB spectrum.} However, an interest of the theory
developed in
\cite{delong,decourt} is that $B$ is not a free parameter (unlike many other
cosmological models that depend on one or several free parameters) but is
fixed by physical considerations. Therefore, it can be interpreted as a sort of
fundamental constant with the value $B=3.53\times 10^{-3}$, which is of the 
order of the inverse  of the famous number $123$ occuring in the so-called
cosmological constant problem.\footnote{More precisely,  $B\simeq
1/\ln(\rho_P/\rho_{\Lambda})\simeq 1/[123\ln(10)]$. There is
a conversion factor $\ln(10)$ between decimal and Napierian logarithms.}
Intriguingly, the small but nonzero value of $B$ is related to the nonzero
value of the Planck constant $\hbar$. This suggests that quantum mechanics plays
a role at the cosmological scale in relation to DM and DE.

On the other hand, even if the Logotropic and $\Lambda$CDM models are
close to each other at large (cosmological) scales, they differ at small
(galactic) scales where the $\Lambda$CDM model poses problem. In particular, 
the Logotropic model is able to solve the CDM crisis (cusp problem, missing
satellite problem...). Furthermore, it is able to explain the
universality of the surface density $\Sigma_0=\rho_0 r_h$ of DM halos and can
predict its observed value $\Sigma_0=141 \, M_{\odot}/{\rm pc}^2$
\cite{donato} without arbitrariness \cite{delong,decourt}.

For these reasons, the Logotropic model is a model of cosmological
interest. We have obtained analytical expressions of the statefinders and shown
that they slightly differ from the values of the  $\Lambda$CDM
model. The quantity of most interest seems to  be the parameter $s$ whose
predicted current value, $s_0=-B/(B+1)=-0.003518$, is directly related to the
fundamental constant $B=3.53\times 10^{-3}$ of the Logotropic model
independently of any other parameter.

Finally, an interesting aspect of our paper is to demonstrate
explicitly that two cosmological models can be indistinguishable at large scales
at the present time
while they have a completely different evolution in the future since the
Logotropic model leads to a phantom evolution (the energy density increases with
the scale factor) unlike the $\Lambda$CDM model (the energy density tends to a
constant). This result is interesting on a cosmological, physical and even
philosophical point of view.

\section*{Acknowledgments}
S.K. gratefully acknowledges the support from SERB-DST project No. EMR/2016/000258.

\appendix

\section{Generalized thermodynamics and effective temperature}
\label{sec_gtet}

In this Appendix, we show that the Logotropic equation of state (\ref{ldf1}) can
be related
to a notion of (effective) generalized thermodynamics.\footnote{The analogy
with generalized thermodynamics was mentioned in Ref. \cite{delong} and
is here systematically developped.} In this approach, the
constant $A$ can be interpreted as a generalized temperature called the
Logotropic temperature \cite{delong,decourt}. Generalized
thermodynamics was introduced by Tsallis \cite{tsallis} and developed by
numerous authors. The underlying idea of generalized
thermodynamics is to notice that many results obtained
with the Boltzmann entropy
can be extended to more general entropic functionals. The formalism  of
generalized
thermodynamics is
mathematically consistent but the physical interpretation of the generalized
entropy must be discussed in each case. We refer to
\cite{tsallisbook,nfp,entropy}
for recent books and reviews on the subject.

Let us consider a generalized entropy of the form
\begin{equation}
S=-\int C(\rho)\, d{\bf r},
\label{ann1}
\end{equation}
where $C(\rho)$ is a convex function ($C''>0$). Following the fundamental
principle of thermodynamics, the equilibrium state of the system in the
microcanonical
ensemble is obtained by maximizing
the entropy at fixed mass $M=\int \rho\, d{\bf r}$ and energy
$E=\frac{1}{2}\int \rho\Phi\, d{\bf r}$, where $\Phi({\bf r})=\int u(|{\bf
r}-{\bf r}'|)\rho({\bf r}')\, d{\bf r}'$ is the self-consistent mean
field potential ($u(|{\bf r}-{\bf r}'|)$ represents the binary potential of
interaction between the particles which, in the present context, corresponds to
the gravitational interaction). We write the variational problem for the
first variations as
\begin{equation}
\delta S-\beta\delta E-\alpha\delta M=0,
\label{ann2}
\end{equation}
where $\beta=1/T$ and $\alpha$ are Lagrange multipliers that can be interpreted
as an inverse generalized temperature and a  generalized chemical potential.
Performing the variations, we obtain the relation
\begin{equation}
C'(\rho)=-\beta\Phi({\bf r})-\alpha.
\label{ann3}
\end{equation}
This integral equation fully  determines the density $\rho({\bf r})$ since $C'$
is
invertible.  Equation (\ref{ann3}) may be rewritten as $\rho({\bf
r})=F[\beta\Phi({\bf r})+\alpha]$ where $F(x)=(C')^{-1}(-x)$. We note that, at
equilibrium, the density is a function of the potential: $\rho=\rho(\Phi)$.
Taking the derivative of Eq. (\ref{ann3}) with respect to $\rho$, we get
\begin{equation}
\rho'(\Phi)=-\frac{\beta}{C''(\rho)}.
\label{ann3b}
\end{equation}
Equation (\ref{ann3}) determines the equilibrium distribution $\rho({\bf
r})=F[\beta\Phi({\bf r})+\alpha]$ with $F(x)=(C')^{-1}(-x)$ for a given entropy
$C(\rho)$. Inversely,
if the equilibrium distribution  is characterized by a relation
of the form $\rho({\bf
r})=F[\beta\Phi({\bf r})+\alpha]$, the corresponding generalized entropy is
given by
\begin{equation}
C(\rho)=-\int^{\rho}F^{-1}(x)\, dx.
\label{ann3c}
\end{equation}

Taking
the gradient of  Eq. (\ref{ann3}), we get 
\begin{equation}
T\rho C''(\rho)\nabla\rho+\rho\nabla\Phi={\bf 0}.
\label{ann4}
\end{equation}
Comparing this expression with the condition of hydrostatic equilibrium
\begin{equation}
\nabla P+\rho\nabla\Phi={\bf 0},
\label{ann5}
\end{equation}
we obtain
\begin{equation}
P'(\rho)=T\rho C''(\rho),
\label{ann6}
\end{equation}
which can be integrated into 
\begin{equation}
P(\rho)=T\left\lbrack \rho C'(\rho)-C(\rho)\right\rbrack=T\rho^2\left\lbrack
\frac{C(\rho)}{\rho}\right\rbrack'
\label{ann7}
\end{equation}
up to an additive constant. This equation determines the equation of state
$P(\rho)$ associated with the generalized entropy $C(\rho)$. Inversely, for a
given equation of state $P(\rho)$, we find that the generalized entropy is
given by 
\begin{equation}
C(\rho)=\frac{\rho}{T}\int^{\rho} \frac{P(\rho')}{\rho'^2}\,
d\rho'
\label{ann8}
\end{equation}
up to a term of the form $A\rho$, yielding a term proportional to $M$ in
Eq. (\ref{ann1}). Taking the derivative of Eq. (\ref{ann8}), we get
\begin{equation}
C'(\rho)=\frac{P(\rho)}{T\rho}+\frac{1}{T}\int^{\rho}
\frac{P(\rho')}{\rho'^2}\, d\rho'=\frac{1}{T}\int^{\rho}
\frac{P'(\rho')}{\rho'}\, d\rho',
\label{ann8b}
\end{equation}
where we have used an integration by parts to obtain the second equality. Using 
Eq. (\ref{ann8b}), the equilibrium condition (\ref{ann3}) can be rewritten as
\begin{equation}
\int^{\rho}
\frac{P'(\rho')}{\rho'}\, d\rho'=-\Phi-\alpha T.
\label{ann8bb}
\end{equation}
Taking the derivative of Eq. (\ref{ann8bb}) with respect to $\rho$, we get
\begin{equation}
\frac{P'(\rho)}{\rho}=-\frac{1}{\rho'(\Phi)}.
\label{ann8c}
\end{equation}

The generalized free energy is defined by the Legendre
transform
\begin{equation}
F=E-TS.
\label{ann9}
\end{equation}
Using Eq. (\ref{ann8}), we get
\begin{equation}
F=\frac{1}{2}\int\rho\Phi\, d{\bf r}+T\int C(\rho)\, d{\bf
r}=\frac{1}{2}\int\rho\Phi\, d{\bf
r}+\int\rho\int^{\rho}
\frac{P(\rho')}{\rho'^2}\,
d\rho'\, d{\bf r}.
\label{ann10}
\end{equation}
The equilibrium state in the canonical ensemble (in which 
the temperature $T$ is fixed) is obtained by minimizing the free energy at fixed
mass $M=\int \rho\, d{\bf
r}$. The variational problem for the first variations writes
\begin{equation}
\delta F-\mu\delta M=0,
\label{ann11}
\end{equation}
where $\mu$ is a chemical potential. This  leads
again to Eqs. (\ref{ann3}) and
(\ref{ann8bb}) with $\mu=-\alpha T$. The
maximization
of entropy at
fixed mass and energy corresponds to a condition of microcanonical stability
while the minimization of free energy at fixed mass corresponds to a condition
of canonical stability. Although these optimization problems have the same
critical points (cancelling the first order variations), the microcanonical  and
canonical stability of the system (related to the sign of the second
order variations) may differ in the case of ensemble inequivalence. The
condition of canonical stability requires that
\begin{equation}
\delta^2 F=T\int C''(\rho)\frac{(\delta\rho)^2}{2}\, d{\bf
r}+\frac{1}{2}\int\delta\rho\delta\Phi\, d{\bf r}>0
\label{ann11b}
\end{equation}
or, equivalently,
\begin{equation}
\delta^2 F=-\frac{1}{2}\left\lbrace \int
\frac{(\delta\rho)^2}{\rho'(\Phi)}\, d{\bf
r}-\int\delta\rho\delta\Phi\, d{\bf r}\right\rbrace>0
\label{ann11c}
\end{equation}
for all perturbations $\delta\rho$ that conserve mass: $\delta M=0$. On the
other hand, the condition of microcanonical stability requires that the
inequalities of Eqs. (\ref{ann11b}) and (\ref{ann11c}) be satisfied for all
perturbations $\delta\rho$
that conserve mass {\it and} energy at first order: $\delta M=\delta E=0$.
Although
canonical stability always implies microcanonical stability, the converse is
not true in the case of ensemble inequivalence. Ensemble inequivalence may occur
for systems with long-range interactions such
as self-gravitating systems \cite{paddy,cc,campabook}.

We note that the second term of the free energy (\ref{ann10}) can be interpreted
as an
internal energy $U=-TS=\int u(\rho)\, d{\bf r}$. The density of internal energy
$u(\rho)=TC(\rho)=\rho\int^{\rho}
[{P(\rho')}/{\rho'^2}]\,
d\rho'$ satisfies the
first law of thermodynamics $d(u/\rho)=-Pd(1/\rho)$.\footnote{Expanding this
relation, we find that $du=[(P+u)/\rho]d\rho=[h(\rho)/\rho]d\rho$ which is
compatible with Eq.
(\ref{df7}). The difference
between the energy density $\epsilon$ and the density of internal energy $u$
corresponds to the constant of integration in the
expression  $\rho\int^{\rho}
[{P(\rho')}/{\rho'^2}]\,
d\rho'$. This constant of integration gives rise to the rest-mass term $\rho
c^2$ representing DM in the interpretation given in Sec. \ref{sec_df}.}
The density of enthalpy
$h(\rho)=u(\rho)+P(\rho)$ is given by 
$h(\rho)=TC(\rho)+P(\rho)=T\rho C'(\rho)=\rho u'(\rho)$. We note that
$u(\rho)=\int^{\rho}
h(\rho')/\rho'\, d\rho'$ and $P(\rho)=h(\rho)-u(\rho)=\rho u'(\rho)-u(\rho)$. 
We also have
$(h/\rho)'=u''(\rho)=TC''(\rho)=P'(\rho)/\rho$. The last equality
corresponds to $dP=\rho
d(h/\rho)$
which is the Gibbs-Duhem relation. Finally, we note that the condition of
hydrostatic equilibrium (\ref{ann5}) [or Eqs. (\ref{ann3}) and
(\ref{ann8bb})]
is equivalent to the condition of constancy of chemical
potential $h(\rho)/\rho+\Phi({\bf r})=-\alpha T=\mu$ given by Landau and
Lifshitz \cite{ll}.

Let us specifically consider the logarithmic entropy 
\begin{equation}
S=\int\ln\left (\frac{\rho}{\rho_P}\right )\, d{\bf r}
\label{ann12}
\end{equation}
introduced in Ref. \cite{logo}. We have $C(\rho)=-\ln(\rho/\rho_P)$. At
equilibrium, using Eq. (\ref{ann3}), we obtain the distribution
\begin{equation}
\rho({\bf r})=\frac{1}{\beta\Phi({\bf r})+\alpha}.
\label{ann13}
\end{equation}
For the harmonic potential $\Phi({\bf r})=(1/2)\omega_0^2r^2$, it corresponds to
the Lorentzian. For the gravitational potential, it leads to DM halos with a
constant surface density $\Sigma_0$ in agreement with the observations
\cite{delong,decourt}. The equation of state,
given by Eq. (\ref{ann7}), associated with the logarithmic entropy
(\ref{ann12})
is the Logotropic equation of state
\begin{equation}
P(\rho)=T\ln \left (\frac{\rho}{\rho_P}\right ).
\label{ann14}
\end{equation}
The logarithmic free energy is
\begin{equation}
F=\frac{1}{2}\int\rho\Phi\, d{\bf r}-T\int\ln\left
(\frac{\rho}{\rho_P}\right )\, d{\bf r}.
\label{ann15}
\end{equation}
These considerations show that the coefficient $A$ in the Logotropic equation of
state (\ref{ldf1})
can be interpreted as a generalized temperature $T$. This is why we 
call it the Logotropic temperature \cite{delong,decourt}.  As a result, the
universality of $A$ (which
explains the constant values of $\Sigma_0$ and $M_{300}$) may be
interpreted by saying that the Universe is ``isothermal'', except
that isothermality does not refer to a linear equation of state $P=\rho k_B
T/m$ associated with the Boltzmann entropy $S_B=-k_B\int (\rho/m)\ln\rho\, d{\bf
r}$,
but to a Logotropic equation of state (\ref{ann14}) associated with a
logarithmic entropy (\ref{ann12})  in a
generalized thermodynamical framework. If the Logotropic
model \cite{delong,decourt} is
correct, it would be a nice confirmation of the interest of generalized
thermodynamics in physics and astrophysics.

We note that, in the context of generalized thermodynamics, $T$
has usually not the dimension of an ordinary temperature. This is the case only
for the standard Boltzmann entropy. In the case of the Logotropic equation of
state, $T=A$ has the dimension of a
pressure or an energy density.  However, $T$ really plays the role of a
generalized
thermodynamic temperature since it satisfies the fundamental relation [see Eq.
(\ref{ann2})]:
\begin{equation}
\beta=\frac{1}{T}=\frac{\partial S}{\partial E}.
\label{ann16}
\end{equation}
Actually, we can change the definition of the logarithmic entropy so that  $S$
and $T$ really have the dimension of an entropy and a temperature.
We write
\begin{equation}
S=k_B \frac{\rho_{\Lambda}}{m_{\Lambda}}\int\ln\left (\frac{\rho}{\rho_P}\right
)\, d{\bf r},\qquad P(\rho)=\rho_{\Lambda}\frac{k_B T}{m_{\Lambda}}\ln \left
(\frac{\rho}{\rho_P}\right ),\qquad B=\frac{k_B T}{m_{\Lambda}c^2},
\label{las1}
\end{equation}
where the last relation is obtained by comparing the second relation with Eq.
(\ref{add1}). Under that form, we see that $B$ can really be interpreted as a
dimensionless Logotropic temperature. It remains for us to specify the mass
scale $m_{\Lambda}$. It is
natural to take\footnote{We stress that the results of our paper do not depend
on the choice of
$m_{\Lambda}$.}  
\begin{equation}
m_{\Lambda}=\frac{\hbar H_0}{c^2}=1.43\times
10^{-33}\, {\rm eV}/c^2.
\label{las2}
\end{equation}
This mass  scale is often interpreted as the
smallest mass of the bosons predicted by string theory \cite{axiverse} or as
the upper bound on the mass of the graviton \cite{graviton}. It is simply
obtained by equating the Compton wavelength of the particle $\lambda_c=\hbar/mc$
with the Hubble radius $R_H=c/H_0$ (the typical size of the visible Universe).
Since $H_0^2\sim G\epsilon_0/c^2\sim G\rho_{\Lambda}$, alternative expressions
of this mass scale are $m_{\Lambda}\sim
({\hbar}/{c^2})\sqrt{G\rho_{\Lambda}}\sim ({\hbar}/{c^2})\sqrt{\Lambda/8\pi}$,
where $\Lambda=8\pi G\rho_{\Lambda}$ is the cosmological constant. 
The temperature is $k_B T=B m_{\Lambda}
c^2\simeq m_{\Lambda}
c^2/\ln(\rho_P/\rho_{\Lambda})\simeq 1.43\times 10^{-33}\, {\rm
eV}/[123\ln(10)]$. The current value of the logarithmic entropy is
\begin{equation}
S_0/k_B\sim \frac{M_{\Lambda}}{m_{\Lambda}}\ln \left
(\frac{\rho_{\Lambda}}{\rho_P}\right )\sim -\frac{\rho_P}{\rho_{\Lambda}}\ln
\left
(\frac{\rho_P}{\rho_{\Lambda}}\right )\sim -10^{123}\times 123\ln(10),
\label{las3}
\end{equation}
where $M_{\Lambda}\sim \rho_{\Lambda}R_H^3\sim c^3/GH_0=1.04\times
10^{89}\, {\rm eV}/c^2$  is the
typical mass of the visible Universe and we have used the relation
$M_{\Lambda}/m_{\Lambda}=c^5/G\hbar H_0^2\sim \rho_P/\rho_{\Lambda}$ which can
be easily checked. Finally, we note that the current value of the
logarithmic free energy is 
\begin{equation}
F_0=E_0-TS_0\simeq E_0+B M_{\Lambda}c^2\ln\left
(\frac{\rho_P}{\rho_{\Lambda}}\right
)\simeq E_0+M_{\Lambda}c^2,
\label{las4}
\end{equation}
where we have used  $B\simeq 1/\ln(\rho_P/\rho_{\Lambda})$ (see Sec.
\ref{sec_lt}). In the last identity $M_{\Lambda}c^2=1.04\times
10^{89}\, {\rm eV}$ may be interpreted
as the rest mass energy of the Universe. 

{\it Remark:} We note that the logarithmic entropy 
is negative (because $\rho<\rho_P$). Actually, we could define the entropy
with the opposite sign but, in that
case, $T$ would become negative in order to ensure the condition $B>0$ (this
condition is necessary to match the observations \cite{delong}). With this new
convention:
\begin{equation}
S=-k_B \frac{\rho_{\Lambda}}{m_{\Lambda}}\int\ln\left (\frac{\rho}{\rho_P}\right
)\, d{\bf r},\qquad P(\rho)=-\rho_{\Lambda}\frac{k_B T}{m_{\Lambda}}\ln \left
(\frac{\rho}{\rho_P}\right ),\qquad B=-\frac{k_B T}{m_{\Lambda}c^2}.
\label{las5}
\end{equation}
Therefore, the concept of negative temperature ($T<0$), which is required in
order to have a positive logarithmic entropy ($S>0$), may explain in a
relatively natural manner why the pressure of the DF (which is responsible for
the accelerating expansion of the Universe) is negative ($P=\rho_{\Lambda}({k_B
T}/{m_{\Lambda}})\ln \left
({\rho_P}/{\rho}\right )<0$). These results will have to
be discussed further in future works. It would also be
interesting to
investigate a possible connection between the logarithmic entropy and  the
holographic principle \cite{bousso}.

\end{document}